\documentclass[aap,seceqn,MSNbibl,citesort,dvips]{arximspdf}
\usepackage{mathrsfs}

\doi{10.1214/09-AAP669}
\volume{20}
\issue{5}
\pubyear{2010}
\firstpage{1697}
\lastpage{1728}

\makeatletter

\newtheorem{theorem}{Theorem}[section]
\newproclaim{defn}[theorem]{Definition}
\newtheorem{prop}[theorem]{Proposition}
\newproclaim{rem}[theorem]{Remark}
\newproclaim{exa}[theorem]{Example}

\newcommand{\cS}{{}^\mathsf{c}\hspace*{-1pt}S}
\newcommand{\cMM}{{}^\mathsf{c}\hspace*{-1pt}[M, M]}

\newcommand{\llbracket}{[\![}
\newcommand{\rrbracket}{]\!]}

\def\bsuffix #1{#1}

\makeatother

\begin{document}
\begin{frontmatter}

\title{Num\'{e}raire-invariant preferences in financial modeling\thanksref{TZZ}}
\runtitle{Num\'{e}raire-invariant preferences in financial modeling}

\pdftitle{Numeraire-invariant preferences in financial modeling}

\thankstext{TZZ}{Supported by NSF Grant DMS-09-08461.}

\begin{aug}
\author[A]{\fnms{Constantinos} \snm{Kardaras}\corref{}\ead[label=e1]{kardaras@bu.edu}}
\runauthor{C. Kardaras}
\affiliation{Boston University}
\address[A]{Department of Mathematics and Statistics\\
Boston University\\
111 Cummington Street\\
Boston, Massachusetts 02215\\
USA\\
\printead{e1}} 
\end{aug}

\received{\smonth{3} \syear{2009}}
\revised{\smonth{11} \syear{2009}}

%
\begin{abstract}
We provide an axiomatic foundation for the representation of
num\'{e}raire-invariant preferences of economic agents acting in a financial
market. In a static environment, the simple axioms turn out to be
equivalent to the following choice rule: the agent prefers one outcome
over another if and only if the expected (under the agent's subjective
probability) relative rate of return of the latter outcome with respect
to the former is nonpositive. With the addition of a transitivity
requirement, this last preference relation has an extension that can be
numerically represented by expected logarithmic utility. We also treat
the case of a dynamic environment where consumption streams are the
objects of choice. There, a novel result concerning a canonical
representation of unit-mass optional measures enables us to explicitly
solve the investment--consumption problem by separating the two aspects
of investment and consumption. Finally, we give an application to the
problem of optimal num\'{e}raire investment with a random time-horizon.
\end{abstract}

%

\pdfkeywords{60G07, 91B08, 91B16, 91B28, Preferences,
choice rules, numeraire-invariance, optional measures,
investment--consumption problem,
random time-horizon utility maximization}

\end{frontmatter}

\setcounter{section}{-1}

\section{Introduction}

Within the class of expected utility maximization problems in economic
theory, the special case of maximizing expected \textit{logarithmic}
utility has undoubtedly attracted considerable attention. The major
reason for its celebrity is the computational advantage it offers: the
use of the logarithmic function allows for explicit solutions of the
optimal investment--consumption problem in general semimartingale models
(see \cite{MR1970286}). Furthermore, in many diverse applications,
optimal portfolios stemming from expected log-utility maximization are
crucial. We mention, for example, the problem of quantifying the
additional utility of a trader using insider information (see \cite
{MR2223957} and the references therein), as well as the use of the
log-optimal portfolios as benchmarks in financial theory, as is
presented in~\cite{MR2267213}.

The emergence of expected log-utility maximization dates as back as
1738, when Daniel Bernoulli offered a solution to the St. Petersburg
paradox, which can be found in the translated manuscript
\cite{MR0059834}. Bernoulli's use of the logarithmic (and, indeed, of any
other increasing and concave) utility function was ad-hoc and lacked
any axiomatization based on rational agent's choice behavior. In the
context of financial choice, \cite{Williams36} seems to be the first
work that has proposed maximizing growth as a reasonable optimization
criterion, which is exactly consistent with expected log-utility
optimization. After Kelly's information-theoretical justification of
using growth-optimal strategies in \cite{MR0090494}, there had been
further attempts to justify maximizing expected log-utility, for
example, in \cite{RePEcucpjpolecv67y1959p144}. Along came heavy
criticism by distinguished scholars, notably by Samuelson (see
\cite{MR0295739} and \cite{RePEceeejbfinav3y1979i4p305307}).
However, the interest in log-optimality has not ceased, and is even
growing. Statistical or behavioral tests do not seem to uniformly favor
one side or the other; for example, Long's work \cite{LONG},
which has inspired some of the recent development, fails to answer with
statistical significance the question whether the log-optimal portfolio
coincides with the market portfolio.

In spite of all the debate that has prolonged over the years, there has
been no attempt in the realms of the theory of choice to investigate
the exact behavioral axioms that, when imposed, would explain the cases
where agents act as if they are maximizing expected logarithmic utility
under a subjective probability measure. Of course, there has been
immense work on axiomatizing agent's preferences, with \cite{MR2316805}
being the first example where axioms were imposed ensuring that agents
act like they are maximizing expected utility over lotteries with a
known statistical nature of the uncertain environment. Savage's work
\cite{MR0348870} provided an axiomatic framework where both the
statistical views and the utility function came as a byproduct. Since
then, there have been numerous successful efforts in relaxing in some
direction the axioms in order to explain agents' behavior in more
depth. In all these works, the representation of preferences via
utilities of logarithmic shape does not appear to have any form of
significance. Naturally, there are descriptive characterizations
aplenty; for example, one could argue that agents that act consistently
with maximizing expected log-utility have constant, and equal to unit,
relative risk aversion. However, a normative characterization seems to
be absent in the literature.

The purpose of this paper is to address the aforementioned issue.
Certain axioms are proposed on the choice of agents amongst random
outcomes that result in the following preference representation: agents
act as if they were making choices based on the \textit{expected relative
rate of return} of an outcome with respect to some alternative based on
a \textit{subjective} probability measure. In particular, an outcome will
be preferred over another if the expected relative rate of return of
the latter with respect to the former outcome is nonpositive. Choices
based on the previous rule are closely connected to preferences
stemming from a numerical representation of expected logarithmic
utility, as can be seen using first-order conditions for optimality.
Actually, we shall discuss how one can extend preferences based on
expected relative rates of return to preferences that have a numerical
representation of expected logarithmic utility, by imposing an extra
transitivity axiom. However, working with expected relative rate of
return is far more appealing, as the agent is not forced to express a
preference between all pairs of alternatives; in other words, the
preference relation will not be complete. The agent is only required to
be able to make choices from certain convex \textit{bundle sets}; in this
respect, we take a more behavioral route in formulating preferences via
\textit{choice rules}.

The key axiom that is imposed to ensure that an agent makes choices
according to the intuitive way described above is the
\textit{num\'{e}raire invariance} of preferences---this simply means that the agent's
comparison of one outcome to another does not depend on the units that
these outcomes are denominated. This is clearly necessary if we are
using expected relative rate of return as a means of comparison, as
relative rates of return do not depend on the denomination.
Furthermore, preferences with expected logarithmic utility
representation are also num\'{e}raire invariant, as follows from the simple
fact that the logarithmic function transforms multiplication to addition.

We also consider the extension of the preferences in a dynamic
environment where agents make choices over \textit{consumption streams}.
The theory regarding choice is more or less a straightforward extension
of the previous static case; ``subjective probabilities'' are now
defined on a product space of states and time. The novel element is a
decomposition of unit-mass optional measures on the last product space
in two parts: one that has the interpretation of subjective views on
the state space (the interpretation being somewhat loose, since it
might involve density processes that are local martingales instead of
martingales) and another that acts as an agent-specific consumption
clock. This decomposition, a result that sharpens Dol\'{e}ans's
characterization of optional measures, allows for a solution of the
investment--consumption problem for an agent with num\'{e}raire-invariant
preferences that separates the investment and consumption parts of
Merton's problem in a general semimartingale-asset-price setting. A
further application discussed in the text is a solution to the pure
investment log-utility maximization problem with a time-horizon that is
random but not necessarily a stopping time with respect to the agent's
information flow. Such problems have lately been discussed in the
context of credit risk and defaults (see, e.g., \cite{MR2212119}
and \cite{MR2456470}).

From a mathematical point of view, the results of the present paper
concern geometric and topological properties of ${\mathbb{L}^0_+}$
and, in the
dynamic case, of the space of adapted, right-continuous, nonnegative
and nondecreasing processes. The rich structure of the previous very
important spaces is still the subject of scrutinized study (see
\cite{MR1768009,Zit08}); this work contributes to this line of research.

The structure of the paper is simple. Section \ref{sec:static}
contains all the foundational results for the static case, which
includes in particular the axiomatization of num\'{e}raire-invariant
preferences. The dynamic case is treated in Section \ref{sec:dynamic},
where the main focus is on a canonical representation of unit-mass
optional measures and the applications it has for the num\'{e}raire-invariant
investment--consumption problem, as well as for the num\'{e}raire property
under random sampling.

\section{\texorpdfstring{Num\'{e}raire-invariant preferences: The static
case}{Numeraire-invariant preferences: The static case}}
\label{sec:static}

\subsection{Definitions and notation} \label{subsec:probnotanddefn}

Throughout, $\mathbb R_+$ denotes the nonnegative real numbers and
$\mathbb R_{++}$ denotes\vadjust{\goodbreak}
the strictly positive real numbers. For $x \in\mathbb R_+$
and $y \in\mathbb R_{+}$, $x / y$ is defined as usual when $y \in
\mathbb R
_{++}$. When $x \in\mathbb R_{++}$ and $y = 0$, we set $x / y = \infty$.
Finally, if $x = y = 0$, we set $x / y = 1$. This last nonconventional
definition will allow for good bookkeeping in the sequel.

On the probability space $(\Omega, \mathcal{F})$ we consider a family
$\Pi$ of
all probabilities that are equivalent to some baseline probability
$\overline{\mathbb{P}}$. All probabilities in $\Pi$ have the same
sets of zero measure
which we shall be calling \textit{$\Pi$-null}. A set will be called
\textit{$\Pi$-full} if its complement is $\Pi$-null. We write ``$\Pi
$-a.s.'' to mean $\mathbb{P}$-a.s. with respect to any, and then all,
$\mathbb{P}
\in\Pi$. All relationships between random variables are to be
understood in the $\Pi$-a.s. sense: for example, $f \leq g$ means that
$\{f \leq g \}$ is $\Pi$-full. The indicator function of $A \in
\mathcal{F}$ is
denoted by $\mathbb{I}_A$; we use simply $1$ for $\mathbb{I}_\Omega
$. Also,
``$\mathbb{E}_\mathbb{P}$'' denotes expectation under the probability
$\mathbb{P}
\in
\Pi$.

The vector space of equivalence classes of random variables under $\Pi
$-a.s. equality is denoted by $\mathbb{L}^0$. Following standard
practice, we do
not distinguish between a random variable and the equivalence class it
generates. We endow $\mathbb{L}^0$ with the usual metric topology: a sequence
$(f^n)_{n \in\mathbb N}$ in $\mathbb{L}^0$ converges to $f \in
\mathbb{L}^0$ if and only
if for all $\epsilon> 0$ we have $\lim_{n \to\infty}\mathbb
{P}[|f^n - f| > \epsilon] =
0$, where $\mathbb{P}$ is any probability in $\Pi$. Thus, $\mathbb
{L}^0$ becomes a
topological vector space. Whenever we consider a topological property
(e.g., limits or closedness), it will be understood under the
aforementioned metric topology, unless explicitly noted otherwise. A
set $\mathcal{C}\subseteq\mathbb{L}^0$ is called \textit{bounded}
if $\lim_{\ell\to
\infty} ( \sup_{f \in\mathcal{C}} \mathbb{P}[|f| > \ell] ) = 0$
holds for
some, and then for all, $\mathbb{P}\in\Pi$. Furthermore, a set
$\mathcal{C}
\subseteq\mathbb{L}^0$ will be called \textit{convexly compact} if
it is
convex, closed and bounded. The last terminology is borrowed from
\cite{Zit08}, where one can find more information, particularly on
explaining the appellation; convexly compact sets share lots of
properties of convex and compact sets of Euclidean spaces.

We define \mbox{${\mathbb{L}^0_+}:= \{f \in{\mathbb{L}^0_+}\mid f \geq0$,
$\Pi$}-a.s.$\}$ and \mbox{${\mathbb{L}^0_{++}}= \{f \in{\mathbb
{L}^0_+}\mid f > 0$}, $\Pi$-a.s.$\}$. Note that ${\mathbb
{L}^0_{++}}$ is the subset of $\Pi$-a.s.
\textit{strictly} positive random variables and is \textit{not} equal to
${\mathbb{L}^0_+}
\setminus\{0 \}$. A set $\mathcal{C}\subseteq{\mathbb{L}^0_+}$ is
called \textit{solid}
if the conditions $0 \leq f \leq g$ and $g \in\mathcal{C}$ imply that
$f \in
\mathcal{C}
$ as well. The set $\mathcal{C}\subseteq{\mathbb{L}^0_+}$ will be
called \textit{log-convex} if for all $f \in\mathcal{C}$, all $g \in\mathcal{C}$
and all $\alpha
\in
[0,1]$, the geometric mean $f^\alpha g^{1 - \alpha}$ belongs to
$\mathcal{C}$
as well.

\subsection{Preferences induced by expected relative rates of return}
\label{subsec:inducedpreferences}

In (\ref{eq:relrateLzero}) below and all that follows we are using
the division conventions explained in the first paragraph of Section
\ref{subsec:probnotanddefn}.

Fix $\mathbb{P}\in\Pi$ and set
%
\begin{equation} \label{eq:relrateLzero}
\mathsf{rel}_\mathbb{P}( f | g ) := \mathbb{E}_\mathbb{P}[f / g ] -
1\qquad \mbox{for all } f \in
{\mathbb{L}^0_+}
\mbox{ and } g \in{\mathbb{L}^0_+}.
\end{equation}
In words, $\mathsf{rel}_\mathbb{P}( f | g )$ is the expected, under
$\mathbb{P}$, rate of return
of $f$ in units of $g$; we therefore call $\mathsf{rel}_\mathbb{P}( f
| g )$ the \textit{expected relative rate of return of $f$ with respect to $g$ under
$\mathbb{P}$}. Unless $f = g$, in which case $\mathsf{rel}_\mathbb
{P}( g | f ) = \mathsf{rel}_\mathbb{P}( f | g ) =
0$, it is straightforward to see that the strict inequality $\mathsf
{rel}_\mathbb{P}( g | f ) > - \mathsf{rel}_\mathbb{P}( f | g )$
holds. Also, if $h \in{\mathbb{L}^0_{++}}$, $\mathsf{rel}_\mathbb
{P}( f/h | g/h )
= \mathsf{rel}_\mathbb{P}( f | g )$; the expected relative rate of
return operation is num\'{e}raire-invariant.

For $\mathbb{P}\in\Pi$, the \textit{preference relation}
$\preccurlyeq_\mathbb{P}$ is
defined to be the following binary relation on ${\mathbb{L}^0_+}$:
%
\begin{equation} \label{eq:preferenceviaQ}
\mbox{for } f \in{\mathbb{L}^0_+}\mbox{ and } g \in{\mathbb
{L}^0_+}\qquad f \preccurlyeq_\mathbb{P}g
\quad\Longleftrightarrow\quad\mathsf{rel}_\mathbb{P}( f | g ) \leq0.
\end{equation}
By our division conventions, $f \preccurlyeq_\mathbb{P}g$ holds if
and only if $\{f > 0 \} \subseteq\{g > 0 \}$ and $\mathbb{E}_\mathbb
{P}[f / g \mid g > 0] \leq1$.

Given the preference relation $\preccurlyeq_\mathbb{P}$, the
\textit{strict} preference
relation $\prec_\mathbb{P}$ is defined by requiring that $f \prec
_\mathbb{P}g$ if and only
if $f \preccurlyeq_\mathbb{P}g$ holds and $g \preccurlyeq_\mathbb
{P}f$ fails. It is straightforward
to check that $f \prec_\mathbb{P}g \Longleftrightarrow\mathsf
{rel}_\mathbb{P}( f | g ) < 0$. Note
also that if $f \preccurlyeq_\mathbb{P}g$ and $g \preccurlyeq
_\mathbb{P}f$, then $f = g$, that is,
the equivalence classes for $\preccurlyeq_\mathbb{P}$ are singletons.
[Indeed, if
$\{f \neq g \}$ were not $\Pi$-null, then $0 \leq- \mathsf
{rel}_\mathbb{P}( f | g ) < \mathsf{rel}_\mathbb{P}( g | f ) \leq
0$, which is impossible.]

We list some important properties of the preference relation of
(\ref{eq:preferenceviaQ}).
\begin{theorem} \label{thm:staticproperties}
Fix $\mathbb{P}\in\Pi$ and simply write $\preccurlyeq$ and $\prec$
for the
preference relation $\preccurlyeq_\mathbb{P}$ on ${\mathbb{L}^0_+}$
of (\ref{eq:preferenceviaQ}) and the induced strict preference relation $\prec_\mathbb{P}$. Then:
\begin{enumerate}[(4)]
\item[(1)] $f \preccurlyeq g$ holds if and only if $\{f > 0 \} \subseteq\{g
> 0 \}$ and $(f / g) \mathbb{I}_{\{g > 0 \}} + \mathbb{I}_{\{g = 0 \}}
\preccurlyeq1$.
\item[(2)] If $f \leq g$, then $f \preccurlyeq g$. Furthermore, if $f \leq
g$ and
$\{f = g \}$ is not $\Pi$-full, then $f \prec g$.
\item[(3)] If $h \in{\mathbb{L}^0_+}$, $\{f \in{\mathbb{L}^0_+}\mid f
\preccurlyeq h \}$ is convexly
compact and log-convex, and $\{f \in{\mathbb{L}^0_+}\mid h
\preccurlyeq f \}$ is
convex and log-convex. If actually $h \in{\mathbb{L}^0_{++}}$, $\{f
\in{\mathbb{L}^0_+} \mid h \preccurlyeq f \}$ is further closed.
\item[(4)] If $\mathcal{C}\subseteq\mathbb{L}^0_+$ is convexly compact,
there exists a
unique $\widehat{f}\in\mathcal{C}$ such that $f \preccurlyeq
\widehat{f}$ holds for all $f \in
\mathcal{C}$.
\end{enumerate}
\end{theorem}
\begin{pf}
The proofs of (1) and (2) are straightforward, so we shall focus on
proving (3) and (4). We hold $\mathbb{P}\in\Pi$ fixed and drop any
subscripts ``$\mathbb{P}$'' in the sequel.

(3) Call $\mathcal{C}^{h}_{\preccurlyeq}:= \{f \in{\mathbb
{L}^0_+}\mid f \preccurlyeq h \}$.
From the definition (\ref{eq:relrateLzero}) of $\mathsf{rel}$, it is
clear that $\mathcal{C}^{h}_{\preccurlyeq}$ is convex. Let $(f^n)_{n
\in\mathbb N}$ be a
sequence in $\mathcal{C}^{h}_{\preccurlyeq}$ such that $\lim_{n \to
\infty}f^n = f$. Since $f^n \preccurlyeq h$
for all $n \in\mathbb N$, property (1) implies that $\{h = 0 \}
\subseteq\{f^n = 0 \}$ for all $n \in\mathbb N$. Then, $\{h = 0 \}
\subseteq\bigcap_{n \in\mathbb N} \{f^n = 0 \} \subseteq\{f = 0 \}
$. An application of Fatou's lemma gives $\mathbb{E}[ f / h \mid h
> 0 ] \leq\mathop{\lim\inf}_{n \to\infty}\mathbb{E}[ f^n / h
\mid h > 0 ] \leq1$
which, in view of $\{f > 0 \} \subseteq\{h > 0 \}$, is equivalent to
$\mathsf{rel}( f | h ) \leq0$. Therefore, $\mathcal
{C}^{h}_{\preccurlyeq}$ is closed. Now, $\mathbb{E}[ f /
h \mid h > 0] \leq1$ for all $f \in\mathcal{C}^{h}_{\preccurlyeq}$
gives $\sup_{f \in
\mathcal{C}^{h}_{\preccurlyeq}} \mathbb{P}[f / h > \ell\mid h > 0]
\leq1 / \ell$ for all
$\ell
\in\mathbb R_+$. In other words, $ \{ f \mathbb{I}_{\{h > 0 \}} \mid f
\in\mathcal{C}^{h}_{\preccurlyeq}\} \subseteq{\mathbb{L}^0_+}$ is
bounded. Since $f = f \mathbb{I}
_{\{h > 0 \}}$ holds for all $f \in\mathcal{C}^{h}_{\preccurlyeq}$,
we get that $\mathcal{C}^{h}_{\preccurlyeq}$ is
bounded. We have therefore established the convex compactness of
$\mathcal{C}^{h}_{\preccurlyeq}$. It remains to establish
log-convexity, which is an easy
application of H\"{o}lder's inequality: for $f \in\mathcal
{C}^{h}_{\preccurlyeq}$, $g \in
\mathcal{C}^{h}_{\preccurlyeq}$ and $\alpha\in[0,1]$,
\begin{eqnarray*}
\mathbb{E}\biggl[ \frac{f^\alpha g^{1 - \alpha}}{h} \Bigm| h > 0\biggr]
&=& \mathbb{E}\biggl[ \biggl( \frac{f}{h} \biggr)^\alpha\biggl( \frac{g}{h}
\biggr)^{1-\alpha} \Bigm| h > 0 \biggr] \\
&\leq&
\biggl( \mathbb{E}\biggl[ \frac{f}{h} \Bigm| h > 0 \biggr] \biggr)^{\alpha} \biggl( \mathbb{E}
\biggl[\frac{g}{h} \Bigm| h > 0 \biggr] \biggr)^{1 - \alpha} \leq1,
\end{eqnarray*}
which shows that $(f^\alpha g^{1 - \alpha}) \in\mathcal
{C}^{h}_{\preccurlyeq}$.

Continuing, fix $h \in{\mathbb{L}^0_+}$ and let $\mathcal
{C}^{h}_{\succcurlyeq}:= \{f \in{\mathbb{L}^0_+} \mid h \preccurlyeq
f \}$. The convexity of $\mathcal{C}^{h}_{\succcurlyeq}$ follows from the
definition of $\mathsf{rel}$ and the convexity of the mapping $\mathbb R_+
\ni
x \mapsto1 / x \in\mathbb R_+ \cup\{\infty \}$. Also, log-convexity of
$\mathcal{C}^{h}_{\succcurlyeq}$ follows similarly as log-convexity
of $\mathcal{C}^{h}_{\preccurlyeq}$. If,
furthermore, $h \in{\mathbb{L}^0_{++}}$, closedness of $\mathcal
{C}^{h}_{\succcurlyeq}$ follows directly by
noticing that $\mathcal{C}^{h}_{\succcurlyeq}= \{ f \in{\mathbb
{L}^0_{++}}\mid(1 / f) \in\mathcal{C}
^{1/h}_{\preccurlyeq} \}$ and that $\mathcal{C}^{1/h}_{\preccurlyeq
}$ is closed.

(4) We shall be assuming throughout that $\mathcal{C}\neq
\{0 \}$; otherwise, trivially, $\widehat{f}= 0$.

We begin by showing there exists $g \in\mathcal{C}$ such that $\{f >
0 \}
\subseteq\{g > 0 \}$ holds for all $f \in\mathcal{C}$. Indeed, let
$p :=
\sup\{\mathbb{P}[f > 0] \mid f \in\mathcal{C} \} > 0$. Using the
convexity and
closedness of $\mathcal{C}$, a standard exhaustion argument shows that there
exists $g \in\mathcal{C}$ such that $\mathbb{P}[g > 0] = p$. If $\{f
> 0 \} \cap
\{g = 0 \}$ were not $\Pi$-null for some $f \in\mathcal{C}$, then,
with $h =
(f + g) / 2 \in\mathcal{C}$, we have $\mathbb{P}[h > 0] = \mathbb
{P}[g > 0] + \mathbb{P}[
\{f > 0 \} \cap\{g = 0 \}] > p$, which is impossible.

We claim that, in order to show (4), we may assume that $\mathcal
{C}\cap{\mathbb{L}^0_{++}}
\neq\varnothing$. Indeed, with $g \in\mathcal{C}$ as above, let
${\widetilde{\mathcal{C}}}:= \{f + \mathbb{I}_{\{ g = 0 \}} \mid f
\in\mathcal{C} \}$. It is straightforward that
${\widetilde{\mathcal{C}}}
$ is convexly compact, as well as that ${\widetilde{\mathcal{C}}}\cap
{\mathbb{L}^0_{++}}\neq\varnothing$.
Furthermore, $f \preccurlyeq\widehat{f}$ holds for all $f \in
\mathcal{C}$ if and only if
${\widetilde{f}}\preccurlyeq\widehat{f}+ \mathbb{I}_{\{ g = 0\}}$
holds for all ${\widetilde{f}}\in{\widetilde{\mathcal{C}}}$.
Therefore, changing from $\mathcal{C}$ to ${\widetilde{\mathcal
{C}}}$ if necessary, we may assume that
$\mathcal{C}\cap{\mathbb{L}^0_{++}}\neq\varnothing$.

Since we can assume that $\mathcal{C}\cap{\mathbb{L}^0_{++}}\neq
\varnothing$, we may
additionally assume that
$1 \in\mathcal{C}$. Indeed, otherwise, we consider $\widetilde
{\mathcal{C}} := (1 / g)
\mathcal{C}$ for some $g \in\mathcal{C}\cap\mathbb{L}^0_{++}$.
Then, $1 \in\widetilde{\mathcal{C}}$
and $\widetilde{\mathcal{C}}$ is still convexly compact. Furthermore, $f
\preccurlyeq
\widetilde{f}$ holds for $f \in\widetilde{\mathcal{C}}$, then
$\widehat{f}:= g
\widetilde{f} \in\mathcal{C}$ satisfies $f \preccurlyeq\widehat
{f}$ for all $f \in\mathcal{C}
$ by
the num\'{e}raire-invariance property (1).

In the sequel, assume that $1 \in\mathcal{C}$ and that $\mathcal{C}$
is convexly
compact. We claim that we can further assume without loss of generality
that $\mathcal{C}$ is solid. Indeed, let $\mathcal{C}'$ be the \textit{solid hull} of
$\mathcal{C}$,
that is, $\mathcal{C}' := \{f \in\mathbb{L}^0_{+} \mid0 \leq f \leq
h \mbox{ holds for some } h \in\mathcal{C} \}$. Then, it is
straightforward that $1 \in
\mathcal{C}
'$, as well as that $\mathcal{C}'$ is still convex and bounded. It is
also true
that $\mathcal{C}'$ is still closed. (To see the last fact, pick a
$\mathcal{C}'$-valued
sequence $(f^n)_{n \in\mathbb N}$ that converges $\mathbb{P}$-a.s. to $f
\in
\mathbb{L}^0_{+}$. Let $(h^n)_{n \in\mathbb N}$ be a $\mathcal
{C}$-valued sequence with
$f^n \leq h^n$ for all $n \in\mathbb N$. By Lemma A.1 from
\cite{MR1304434}, we can extract a sequence $({\widetilde{h}}^n)_{n \in
\mathbb N}$ such
that, for each $n \in\mathbb N$, ${\widetilde{h}}^n$ is a convex
combination of
$h^n, h^{n+1}, \ldots,$ and such that $h := \lim_{n \to\infty
}{\widetilde{h}}^n$ exists. Of
course, $h \in\mathcal{C}$ and it is easy to see that $f \leq h$. We then
conclude that $f \in\mathcal{C}'$.) Suppose that there exists
$\widehat{f}\in\mathcal{C}'$
such that $f \preccurlyeq\widehat{f}$ holds for all $f \in\mathcal
{C}'$. Then, $\widehat{f}
\in
\mathcal{C}$ (since $\widehat{f}$ has to be a \textit{maximal} element
of $\mathcal{C}'$ with
respect to the order structure of $\mathbb{L}^0$), and that $f
\preccurlyeq\widehat{f}$
holds for all $f \in\mathcal{C}$ (simply because $\mathcal
{C}\subseteq\mathcal{C}'$).

To recapitulate, in the course of the proof of $(4)$, we shall be
assuming without loss of generality that $\mathcal{C}\subseteq\mathbb
{L}^0_{+}$ is
solid, convexly compact, as well as that $1 \in\mathcal{C}$.

For all $n \in\mathbb N$, let $\mathcal{C}^n := \{f \in\mathcal
{C}\mid f \leq n \}$, which is convexly compact and satisfies
$\mathcal{C}^n \subseteq\mathcal{C}$.
Consider the following optimization problem:
%
\begin{equation} \label{eq:log-optimalprob}
\mbox{find } f_*^n \in\mathcal{C}^n \qquad\mbox{such that } \mathbb
{E}[\log(f_*^n)] =
\sup_{f \in\mathcal{C}^n} \mathbb{E}[\log(f)].
\end{equation}
The fact that $1 \in\mathcal{C}^n$ implies that the value of the
above problem
is not $- \infty$. Further, since $f \leq n$ for all $f \in\mathcal
{C}^n$, one
can use of Lemma A.1 from \cite{MR1304434} in conjunction with the
inverse Fatou's lemma and obtain the existence of the optimizer $f_*^n$
of (\ref{eq:log-optimalprob}). For all $f \in\mathcal{C}^n$ and
$\epsilon
\in\ ]0, 1/2]$, one has
%
\begin{eqnarray} \label{eq:finitediff}
\mathbb{E}[\Delta_\epsilon(f \mid f_*^n ) ] \leq0\hspace*{80pt}\nonumber\\[-8pt]\\[-8pt]
\eqntext{\mbox{where }
\Delta_\epsilon(f \mid f_*^n ) := \dfrac{\log((1 - \epsilon) f_*^n
+ \epsilon f ) - \log(f_*^n )}{\epsilon}.}
\end{eqnarray}
Fatou's lemma will be used on (\ref{eq:finitediff}) as $\epsilon
\downarrow0$. For this, observe that $\Delta_\epsilon(f \mid f_*^n)
\geq0$ on the event $\{f > f_*^n \}$. Also, the inequality $\log(y) -
\log(x) \leq(y - x) / x$, valid for $0 < x < y$, gives that, on $\{f
\leq f_*^n \}$, the following lower bound holds (remember that
$\epsilon\leq1 /2$):
\[
\Delta_\epsilon(f \mid f_*^n) \geq- \frac{f^n_* - f}{f^n_* -
\epsilon(f^n_* - f)} \geq- \frac{f^n_* - f}{f^n_* - (f^n_* - f) / 2}
= - 2 \frac{f^n_* - f}{f^n_* + f} \geq- 2.
\]
Using Fatou's Lemma on (\ref{eq:finitediff}) gives $\mathbb{E}[(f -
f^n_*) / f^n_* ] \leq0$, or, equivalently, that $f \preccurlyeq
f^n_*$, for
all $f \in\mathcal{C}^n$.

Lemma A.1 from \cite{MR1304434} again gives the existence of a sequence
$(\widehat{f}^n)_{n \in\mathbb N}$ such that each $\widehat{f}^n$ is
a finite
convex combination of $f_*^n, f_*^{n+1},\ldots,$ and $\widehat{f}:=
\lim_{n \to\infty}
\widehat{f}^n$ exists. Since $\mathcal{C}$ is convex, $\widehat{f}^n
\in\mathcal{C}$ for all $n
\in
\mathbb N$; therefore, since $\mathcal{C}$ is closed, $\widehat{f}\in
\mathcal{C}$ as well. Fix
$n \in\mathbb N$ and some $f \in\mathcal{C}^n$. For all $k \in
\mathbb N$ with
$k \geq n$, we have \mbox{$f \in\mathcal{C}^k$}. Therefore, $f \preccurlyeq
f^k_*$, for all
$k \geq n$. Since $\widehat{f}^n$ is a finite convex combination of $f_*^n,
f_*^{n+1},\ldots,$ by part (3) of Theorem \ref{thm:staticproperties}
which we already established, we have $f \preccurlyeq\widehat{f}^n$,
that is,
$\mathbb{E}[f / \widehat{f}^n] \leq1$. Then, Fatou's lemma implies
that for all
$f \in\bigcup_{k \in\mathbb N} \mathcal{C}^k$ one has $\mathbb
{E}[f / \widehat{f}]
\leq
1$. The extension of the last inequality to all $f \in\mathcal{C}$
follows from
the solidity of $\mathcal{C}$ by an application of the monotone
convergence theorem.
\end{pf}

Our main point will be to give certain axioms on a preference relation
$\preccurlyeq$ on ${\mathbb{L}^0_+}$ that will imply the
representation given by
(\ref{eq:preferenceviaQ}) for some ``subjective'' probability
$\mathbb{P}\in
\Pi$. This will eventually be achieved in Theorem \ref{thm:converse},
and the properties obtained in Theorem \ref{thm:staticproperties}
above will serve as guidelines. Before that, we slightly digress in
order to better understand the preference relation given by (\ref{eq:preferenceviaQ}), as well as to discuss a class of subsets of
${\mathbb{L}^0_+}$
with a special structure that will prove important.

\subsection[On the relation $\preccurlyeq_\mathbb{P}$ of (1.2)]{On the
relation $\preccurlyeq_\mathbb{P}$ of
(\protect\ref{eq:preferenceviaQ})}
\label{subsec:relationproperties}

For the purposes of Section \ref{subsec:relationproperties}, fix
$\mathbb{P}\in\Pi$ and let $\preccurlyeq$ denote the binary
relation of
(\ref{eq:preferenceviaQ}), dropping the subscript ``$\mathbb{P}$'' from
$\preccurlyeq_\mathbb{P}$. We also simply use ``$\mathsf{rel}$'' to
denote ``$\mathsf{rel}
_\mathbb{P}$'' and ``$\mathbb{E}$'' to denote expectation under
$\mathbb{P}$. Also,
throughout Section \ref{subsec:relationproperties}, we tacitly
preclude the uninteresting case where $\mathbb{L}_+^0$ is isomorphic
to the
nonnegative real line, that is, when $\mathcal{F}$ is trivial modulo
$\Pi$.

As shall soon be revealed, the relation $\preccurlyeq$ fails to
satisfy the
fundamental tenets of a \textit{rational} preference relation, namely,
completeness and transitivity. We shall try nevertheless to argue that
this failure is natural in the present setting.

\subsubsection{Quasi-convexity}
The convexity of the upper-contour set $\{f \in{\mathbb{L}^0_+}\mid h
\preccurlyeq f \}$, where $h \in{\mathbb{L}^0_+}$, makes
$\preccurlyeq$ a so-called \textit{quasi-convex} preference relation. If $\preccurlyeq$ were complete, the
lower-contour sets $\{f \in{\mathbb{L}^0_+}\mid f \preccurlyeq h \}$
would fail to
be convex in general. However, lower-contour sets \textit{are} convex,
according to property (3) of Theorem \ref{thm:preferences}---this
already points out that $\preccurlyeq$ is not complete. The convexity of
$\{f \in{\mathbb{L}^0_+}\mid f \preccurlyeq h \}$ is natural when one
recalls the
definition of the preference relation: if both $f \in{\mathbb
{L}^0_+}$ and $g \in
{\mathbb{L}^0_+}$ have nonpositive expected relative rate of return
with respect to
$h$, so does any convex combination of $f$ and $g$.

\subsubsection{The relation $\preccurlyeq$ is not complete} Pick $A
\in\mathcal{F}$
with $0 < \mathbb{P}[A] < 1$. With $f = \mathbb{I}_{\Omega\setminus
A}$ and $g
= \mathbb{I}_{A}$, we have $\mathsf{rel}( f | g ) = \infty= \mathsf
{rel}( g | f )$; therefore,
neither $f \preccurlyeq g$ nor $g \preccurlyeq f$ holds. One can find more
interesting examples involving elements of ${\mathbb{L}^0_{++}}$. Let
$p := \mathbb{P}
[A]$, $f := (1 / p) \mathbb{I}_A + (1 - p) \mathbb{I}_{\Omega
\setminus A}$
and $g := 1$. Then, $\mathsf{rel}( f | g ) = (1 - p)^2 > 0$ and
$\mathsf{rel}( g | f ) =
p^2 > 0$, that is, neither $f \preccurlyeq g$ nor $g \preccurlyeq f$ holds.

The relation $\preccurlyeq$ is really too strong: $f \preccurlyeq g$
implies that
$g$ is preferred over \textit{any} convex combination of $f$ and $g$.
More precisely, statement (3) of Theorem \ref{thm:staticproperties}
implies that, if $f \preccurlyeq g$ then, for all $\alpha\in[0,1]$ and
$\beta\in[0,1]$ with $\alpha\leq\beta$, we have $(1 - \alpha) f +
\alpha g \preccurlyeq(1 - \beta) f + \beta g$. A pair of $f \in
{\mathbb{L}^0_+}$
and $g \in{\mathbb{L}^0_+}$ will be comparable if and only if one of
$f \in{\mathbb{L}^0_+}$
or $g \in{\mathbb{L}^0_+}$ is preferable over the whole set $\mathsf
{conv}(f, g) :=
\{(1 - \alpha) f + \alpha g \mid\alpha\in[0,1] \}$. The equivalent of
the completeness property here is the following: if $f \in{\mathbb
{L}^0_+}$ and $g
\in{\mathbb{L}^0_+}$, there exists $h \in\mathsf{conv}(f, g)$ that
dominates all elements
in $\mathsf{conv}(f, g)$. In both examples that were given above ($f =
\mathbb{I}
_{\Omega\setminus A}$ and $g = \mathbb{I}_{A}$, as well as $f = (1 / p)
\mathbb{I}_A + (1 - p) \mathbb{I}_{\Omega\setminus A}$ and $g = 1$),
one can
actually check that $h = (1 - p) f + p g$.

\subsubsection{The relation $\preccurlyeq$ is not transitive}
\label{subsubsec:nottransitive}

Pick $A \in\mathcal{F}$ with $0 < \mathbb{P}[A] < 1$. With $p :=
\mathbb{P}[A]$, let $f
:= (1 / p) \mathbb{I}_{A}$, $g := 1$ and $h := ( 2 p / (1 + p)
) \mathbb{I}_A + 2 \mathbb{I}_{\Omega\setminus A}$. It is straightforward
to check that $\mathsf{rel}( f | g ) = 0$, $\mathsf{rel}( g | h ) =
0$, as well as $\mathsf{rel}( f | h ) = (1 - p) / ( 2 p ) > 0$. In
other words, we have $f \preccurlyeq g$
and $g \preccurlyeq h$, but $f \preccurlyeq h$ fails.

Whereas failure of completeness of preference relations is not
considered dramatic, and is indeed welcome in certain cases,
transitivity is a more or less unquestionable requirement. The reason
for its failure in the present context does not have to do with
irrationality of agents making choices according to $\preccurlyeq$. Recall
that $f \preccurlyeq g$ and $g \preccurlyeq h$ mean that $g$ is the
best choice
from the set $\mathsf{conv}(f,g)$ and $h$ the is best choice amongst
$\mathsf{conv}
(g,h)$. However, when an agent is presented with the set of
alternatives $\mathsf{conv}(f,h)$, some strict convex combination of
$f$ and
$h$ might be preferable to $h$, especially when $f$ pays off
considerably better on an event where $h$ does not.

Although $f \preccurlyeq h$ fails in the example above, one expects
that $h
\preccurlyeq f$ fails as well, and this is indeed the case. In general, even
though transitivity does not hold, we have a weaker ``chain'' property
holding.\vspace*{1pt} For $n \in\mathbb N$, let $f^0, \ldots, f^n$ be elements of
${\mathbb{L}^0_+}$ satisfying $f^{i-1} \preccurlyeq f^i$ for $i \in\{
1, \ldots, n \}$
and $f^0 = f^n$. Then, actually, $f^i = f^0$ holds for all $i \in\{1,
\ldots, n \}$. Indeed, let $\phi^i := f^{i - 1} / f^i$ for $i \in
\{1, \ldots, n \}$. We wish to show that $\phi^i = 1$ for all $i
\in
\{1, \ldots, n \}$. Suppose the contrary. Since $\mathbb{E}[\phi^i]
\leq
1$ holds for all $i \in\{1, \ldots, n \}$, the strict convexity of
the mapping $\mathbb R^n_{++} \ni(x^1, \ldots, x^n) \mapsto\prod_{i=1}^n
(1 / x^i)$, combined with the fact that $\mathbb{P}[\phi^i = 1] < 1$ holds
for some $i \in\{1, \ldots, n \}$ and a use of Jensen's inequality
gives $\mathbb{E}[ \prod_{i=1}^n (1 / \phi^i) ] > 1$. However,
$\prod_{i=1}^n (1 / \phi^i) = 1$, which is a contradiction.

\subsubsection{The relation $\preccurlyeq$ does not respect addition} Pick
$A \in\mathcal{F}$ such that $0 < \mathbb{P}[A] \leq1/2$. With $p
:= \mathbb{P}[A]$,
let $f := p^2 \mathbb{I}_A + (1 + p)^2 \mathbb{I}_{\Omega\setminus
A}$ and
$g := p \mathbb{I}_A + (1 + p) \mathbb{I}_{\Omega\setminus A}$.
Observe that
$f = g^2$, $f \neq g$ and $\mathbb{E}[g] = 1$. Then, $\mathsf{rel}( f
| g ) = 0$, so
$f \prec g$. However,
\begin{eqnarray*}
\mathsf{rel}( 1+g | 1+f ) &=& \mathbb{E}\biggl[\frac{g(1 - g)}{1 + g^2} \biggr] =
\frac
{p(1 -
p)}{1 + p^2} p + \frac{(1 + p) (-p)}{1 + (1 + p)^2} (1 - p) \\
&=& \frac{p(1
- p) (p^2 + p - 1)}{(1 + p^2) (1 + (1 + p)^2)} < 0,
\end{eqnarray*}
the last fact following from $p^2 + p - 1 < 0$, which holds in view of
$p \leq1 /2$. Therefore, $1 + g \prec1 + f$. Even though initially
$g$ was preferred to $f$, as soon as the agent is endowed with an extra
unit of account, the choice completely changes. Note that $f$ pays off
very close to zero on $A$; even though $f$ pays off more than $g$ on
$\Omega\setminus A$, a risk-averse agent will prefer $g$. However,
once the risk associated with the outcome $A$ is reduced by the
assurance that a unit of account will be received in any state of the
world, $f$ is preferred.

In fact, regardless of whether $f \prec g$ holds or not, if the event
$\{g < f \}$ is not $\Pi$-null, one can find $h \in{\mathbb{L}^0_+}$
such that $g
+ h \prec f + h$. The proof of this is based on the aforementioned
simple idea: a sufficiently large ``insurance'' $h$ on $\{f \leq g \}$
will make $f + h$ better than $g + h$. Indeed, for $n \in\mathbb N$,
\[
\mathsf{rel}\bigl( g + n g \mathbb{I}_{\{f \leq g \}} | f + n g \mathbb
{I}_{\{f \leq g \}} \bigr)
= \mathbb{E}\biggl[\frac{g - f}{f} \mathbb{I}_{\{g < f \}} \biggr] + \mathbb
{E}\biggl[\frac {g - f}{f + n g} \mathbb{I}_{\{f \leq g \}} \biggr].
\]
The first summand of the right-hand side is strictly negative and the
second one tends to zero as $n \to\infty$ by the monotone convergence
theorem. Therefore, there exists some large enough $N \in\mathbb N$
such that, with $h := N g \mathbb{I}_{\{f \leq g \}}$, \mbox{$\mathsf{rel}(
g + h | f + h ) < 0$}, which completes the argument.

\subsection{Full simplices in ${\mathbb{L}^0_+}$}

We shall describe here a special class of convexly compact sets, which
are the equivalents of simplices with nonempty interior in
finite-dimensional spaces. These sets will turn out to be crucial in
our statement of Theorem \ref{thm:converse} on the axiomatic
definition of num\'{e}raire-invariant preferences. The results
presented here
concern the structure of ${\mathbb{L}^0_+}$; as such, they are of
independent interest.

For $\mathcal{C}\subseteq{\mathbb{L}^0_+}$, define $\mathcal
{C}^{\max}$ to be the subset of $\mathcal{C}$
containing all the \textit{maximal elements} of $\mathcal{C}$, that
is, $f \in
\mathcal{C}^{\max}$ if and only if $f \in\mathcal{C}$ and the
relationships $f \leq g$ and
$g \in\mathcal{C}$ imply that $f = g$.

For a measure $\mu$ on $(\Omega, \mathcal{F})$, we shall write $\mu
\sim\Pi
$ if
$\mu[A] = 0$ holds for all $\Pi$-null $A \in\mathcal{F}$.
\begin{theorem} \label{thm:geomcharofLone}
Let $\mathcal{B}\subseteq{\mathbb{L}^0_+}$. Then, the following
statements are equivalent:
\begin{enumerate}[(3)]
\item[(1)] $\mathcal{B}$ is closed and solid, $\mathcal{B}\cap{\mathbb
{L}^0_{++}}\neq\varnothing$,
$\mathcal{B}^{\max}$
is convex, and $\mathcal{B}= \bigcup_{a \in[0,1]} a \mathcal
{B}^{\max}$.
\item[(2)] For any $\mathbb{P}\in\Pi$, there exists $\widehat{f}=
\widehat{f}(\mathbb{P})
\in
\mathcal{B}\cap{\mathbb{L}^0_{++}}$ such that $\mathcal{B}= \{ f
\in{\mathbb{L}^0_+}\mid f \preccurlyeq_\mathbb{P}
\widehat{f}\}$.
\item[(3)] There exists a $\sigma$-finite measure $\mu\sim\Pi$ such that
$\mathcal{B}= \{f \in{\mathbb{L}^0_+}\mid\int_\Omega f \,d\mu\leq1
\}$.
\end{enumerate}
\end{theorem}
\begin{pf}
We first prove the easy implications $(2) \Rightarrow(3)$ and $(3)
\Rightarrow(1)$; then, $(1) \Rightarrow(2)$ will be tackled.

$(2) \Rightarrow(3)$. Let $\mathbb{P}\in\Pi$ and $\widehat{f}\in
\mathcal{B}
\cap{\mathbb{L}^0_{++}}$ be such that $\mathcal{B}= \{f \in{\mathbb
{L}^0_+}\mid\mathbb{E}_\mathbb{P}[f /
\widehat{f}
] \leq1 \}$. Define $\mu$ via $\mu[A] = \mathbb{E}_\mathbb
{P}[\widehat{f}\mathbb{I}_A]$
for all $A \in\mathcal{F}$. With $A^n := \{ \widehat{f}\leq n \}$
for $n
\in\mathbb N$ we have $\mu[A^n] < \infty$ and $\lim_{n \to\infty
}\mathbb{P}[A^n] = 1$;
therefore, $\mu$ is $\sigma$-finite. Furthermore, $\widehat{f}\in
{\mathbb{L}^0_{++}}$
implies that $\mu\sim\Pi$. The equality $\mathcal{B}= \{f \in
{\mathbb{L}^0_+}
\mid
\int_\Omega f \,d\mu\leq1 \}$ holds by definition.

$(3) \Rightarrow(1)$. Suppose that $\mathcal{B}= \{f \in{\mathbb{L}^0_+}
\mid\int_\Omega f \,d\mu\leq1 \}$ for some $\sigma$-finite
$\mu\sim\Pi$. Closedness of $\mathcal{B}$ follows from Fatou's
lemma and
solidity is obvious from the monotonicity of the Lebesgue integral. As
$\mu$ is $\sigma$-finite, there exists $f \in{\mathbb{L}^0_{++}}$
such that $\int
_\Omega f \,d\mu< \infty$; therefore, $ ( 1 / \int_\Omega f \,d
\mu) f \in\mathcal{B}\cap{\mathbb{L}^0_{++}}$, which shows that
$\mathcal{B}\cap{\mathbb{L}^0_{++}}\neq
\varnothing$. It is straightforward that $\mathcal{B}^{\max}= \{f \in
{\mathbb{L}^0_+}\mid \int_\Omega f \,d\mu= 1 \}$, which implies that
$\mathcal{B}^{\max}$ is convex by
the linearity of Lebesgue integral. For $f \in\mathcal{B}\setminus\{
0 \}$,
set $a := \int_\Omega f \,d\mu\in(0,1]$. Then, $f = a g$, where $g
:= (1 / a) f \in\mathcal{B}^{\max}$. Therefore, $\mathcal{B}=
\bigcup_{a \in[0,1]} a
\mathcal{B}^{\max}$.

$(1) \Rightarrow(2)$. We start by showing that \textit{any
$\mathcal{B}
\subseteq{\mathbb{L}^0_+}$ satisfying the requirements of statement}
(1) \textit{of Theorem
\ref{thm:geomcharofLone} is convexly compact.} Since $\mathcal{B}$ is
closed, only convexity and boundedness of $\mathcal{B}$ have to be established.
We start with convexity. Let $f \in\mathcal{B}$, $g \in\mathcal
{B}$, and $\lambda\in
[0,1]$. We know that there exist $a \in[0,1]$, $b \in[0,1]$, $f' \in
\mathcal{B}^{\max}$ and $g' \in\mathcal{B}^{\max}$ such that $f =
a f'$ and $g = b g'$. Then,
\[
(1 - \lambda) f + \lambda g = \bigl((1 - \lambda) a + \lambda b \bigr)
\biggl(\frac{(1 - \lambda) a}{(1 - \lambda) a + \lambda b} f' + \frac
{\lambda b}{(1 - \lambda) a + \lambda b} g' \biggr),
\]
and the last element belongs to $\mathcal{B}$ due to the fact that
$\mathcal{B}^{\max}$ is
convex and $((1 - \lambda) a + \lambda b ) \in[0,1]$. We have
shown that $\mathcal{B}\subseteq{\mathbb{L}^0_+}$ is convex, solid
and closed. If it were
not bounded, it would follow from Lemma 2.3 in \cite{MR1768009} that
there existed a non-$\Pi$-null $A \in\mathcal{F}$ such that $\{x
\mathbb{I}_A \mid x \in\mathbb R_+ \} \subseteq\mathcal{B}$. But in
that case $\mathcal{B}^{\max}$ would not
contain any element of $\{x \mathbb{I}_A \mid x \in\mathbb R_+ \}$, and
therefore the property $\mathcal{B}= \bigcup_{a \in[0,1]} a \mathcal
{B}^{\max}$ would be
violated. It follows then that $\mathcal{B}$ has to be bounded.

Continuing, fix $\mathbb{P}\in\Pi$. Since $\mathcal{B}$ is convexly
compact and
$\mathcal{B}
\cap{\mathbb{L}^0_{++}}\neq\varnothing$, by Theorem \ref
{thm:staticproperties}(4)
there exists $\widehat{f}\in\mathcal{B}\cap{\mathbb{L}^0_{++}}$
such that $\mathbb{E}_\mathbb{P}[f / \widehat{f}]
\leq1$ holds for all $f \in\mathcal{B}$. Let $\widehat{\mathcal
{B}}:= (1 / \widehat{f}) \mathcal{B}$.
Then, $\widehat{\mathcal{B}}$ also satisfies the requirements of
statement (1) of
Theorem \ref{thm:geomcharofLone}, $1 \in\widehat{\mathcal{B}}^{\max
}$ and $\widehat{\mathcal{B}}
\subseteq\{f \in{\mathbb{L}^0_+}\mid\mathbb{E}_\mathbb{P}[f] \leq
1 \} =: \mathcal{B}^1_\mathbb{P}$.
We shall argue that $\mathcal{B}^1_\mathbb{P}\subseteq\widehat
{\mathcal{B}}$, therefore establishing
that $\widehat{\mathcal{B}}= \mathcal{B}^1_\mathbb{P}$ and
completing the proof. Assume by way of
contradiction that there exists $g \in\mathcal{B}^1_\mathbb
{P}\setminus\widehat{\mathcal{B}}$. Since
$\widehat{\mathcal{B}}$ is closed and solid, it follows that $(g
\wedge M) \notin
\widehat{\mathcal{B}}
$ for large enough $M \in\mathbb R_+$; of course, $(g \wedge M) \in
\mathcal{B}^1_\mathbb{P}
$ also holds, since $\mathcal{B}^1_\mathbb{P}$ is solid. In other
words, we may suppose
that there exists $g \in(\mathcal{B}^1_\mathbb{P}\setminus\widehat
{\mathcal{B}}) \cap{\mathbb{L}^\infty_+}$. Since
$\widehat{\mathcal{B}}= \bigcup_{a \in[0,1]} a \widehat{\mathcal
{B}}^{\max}$, $1 \in\widehat{\mathcal{B}}$,
$\widehat{\mathcal{B}}$
is solid and $g \in{\mathbb{L}^\infty_+}$ does \textit{not} belong to
$\widehat{\mathcal{B}}$, there
exists $a \in(0,1)$ such that $\widetilde{g}:= a g \in\widehat
{\mathcal{B}}^{\max}$. We
shall now establish the following claim (we use $|\cdot|_{\mathbb
{L}^\infty}$ will
denote the usual $\mathbb{L}^\infty$-norm): \textit{$(1 + \epsilon-
\epsilon\widetilde{g})
\in
\widehat{\mathcal{B}}^{\max}$ holds whenever $0 < \epsilon< 1 /
|\widetilde{g}|_{\mathbb{L}^\infty}$}. First
of all, observe that $(1 + \epsilon- \epsilon\widetilde{g}) \in
\mathbb{L}^\infty_+$
whenever $0 < \epsilon< 1 / |\widetilde{g}|_{\mathbb{L}^\infty}$.
Therefore,\vspace*{2pt} since $\widehat{\mathcal{B}}=
\bigcup_{a \in[0,1]} a \widehat{\mathcal{B}}^{\max}$, $1 \in
\widehat{\mathcal{B}}$, and $\widehat{\mathcal{B}}$ is
solid, there exists $b \in\mathbb R_+$ such that $b(1 + \epsilon-
\epsilon\widetilde{g}) \in\widehat{\mathcal{B}}^{\max}$. Since
$\widehat{\mathcal{B}}^{\max}$ is convex and
$\widetilde{g}\in\widehat{\mathcal{B}}^{\max}$, we have
\[
\widehat{\mathcal{B}}^{\max} \ni\biggl(\frac{b \epsilon}{1 + b \epsilon
} \widetilde{g}+ \frac {1}{1 + b \epsilon} b (1 + \epsilon-
\epsilon\widetilde{g}) \biggr) = \frac{b +
b \epsilon}{1 + b \epsilon}.
\]
The last element is a real multiple of $1 \in\widehat{\mathcal
{B}}^{\max}$. Therefore,
$1 = (b + b \epsilon)/(1 + b \epsilon)$, which gives $b = 1$ and
establishes that $(1 + \epsilon- \epsilon\widetilde{g}) \in\widehat
{\mathcal{B}}^{\max}$
whenever $0 < \epsilon< 1 / |\widetilde{g}|_{\mathbb{L}^\infty}$.
But then, with fixed
$\epsilon\in\mathbb R_+$ such that $0 < \epsilon< 1 / |\widetilde
{g}|_{\mathbb{L}^\infty}$, we
have $\mathbb{E}_\mathbb{P}[1 + \epsilon- \epsilon\widetilde{g}] =
1 + \epsilon(1 - a
\mathbb{E}_\mathbb{P}
[g]) > 1$, the last strict inequality holding because $a \in(0,1)$ and
$\mathbb{E}_\mathbb{P}[g] \leq1$. In other words, $(1 + \epsilon-
\epsilon \widetilde{g} ) \notin\mathcal{B}^1_\mathbb{P}$, which is
a contradiction to $\widehat{\mathcal{B}}\subseteq
\mathcal{B}^1_\mathbb{P}
$. We conclude that $\widehat{\mathcal{B}}= \mathcal{B}^1_\mathbb
{P}$, which finishes our argument.
\end{pf}
\begin{defn}
A set $\mathcal{B}\subseteq{\mathbb{L}^0_+}$ satisfying any of the
equivalent statements
of Theorem \ref{thm:geomcharofLone} will be called a \textit{full
simplex in ${\mathbb{L}^0_+}$}.
\end{defn}

The description of a full simplex $\mathcal{B}$ of ${\mathbb{L}^0_+}$
given by (1) in
Theorem \ref{thm:geomcharofLone} is structural. The convex set
$\mathcal{B}^{\max}$ is the ``outer face'' of $\mathcal{B}$, and one
can create the whole set
$\mathcal{B}$ by contracting this face ``inward'' toward zero. This
way one
actually obtains a convexly compact set, though this is not completely
trivial to show. Note that the idea of maximality in ${\mathbb
{L}^0_+}$ was
utilized in order to describe the ``outer face'' $\mathcal{B}^{\max}$
of $\mathcal{B}$.
Theorem \ref{thm:geomcharofLone} shows immediately why
characterizations using topological boundaries would be useless.
Indeed, consider the $\sigma$-finite measure $\mu\sim\Pi$ such that
$\mathcal{B}= \{f \in{\mathbb{L}^0_+}\mid\int_\Omega f \,d\mu\leq1
\}$. Suppose
that $\mathbb{L}^0$ is infinite-dimensional, which is equivalent to the
existence of a sequence $(h^n)_{n \in\mathbb N}$ of elements of
${\mathbb{L}^0_+}$
with $\int_\Omega h^n \,d\mu> 1$ for all $n \in\mathbb N$ and
$\lim_{n \to\infty}
h^n = 0$. Then, the closure of ${\mathbb{L}^0_+}\setminus\mathcal
{B}= \{f \in{\mathbb{L}^0_+} \mid\int_\Omega f \,d\mu> 1 \}$ is
actually equal to ${\mathbb{L}^0_+}$; this is
straightforward once one notices that $f = 0$ belongs in this closure.
Therefore, the topological boundary of the closed set $\mathcal{B}$ is
$\mathcal{B}$ itself.

A preference-theoretic characterization of a full simplex in ${\mathbb
{L}^0_+}$ is
provided in statement (2) of Theorem \ref{thm:geomcharofLone}. For
any probability $\mathbb{P}\in\Pi$, there exists an optimal choice
$\widehat{f}
\in\mathcal{B}$ for $\preccurlyeq_\mathbb{P}$, depending on
$\mathbb{P}$, that makes $\mathcal{B}$ exactly
equal to the lower contour set of $\widehat{f}$.

Statement (3) of Theorem \ref{thm:geomcharofLone} describes a full
simplex $\mathcal{B}$ of ${\mathbb{L}^0_+}$ in a geometric way,
loosely as the intersection
of ${\mathbb{L}^0_+}$ with a half-space. Observe, however, that the
mappings ${\mathbb{L}^0_+}
\ni f \mapsto\int_\Omega f \,d\mu$ for a $\sigma$-finite measure
$\mu
\sim\Pi$ are in general extended-real-valued and not continuous in
${\mathbb{L}^0_+}$. From the perspective of economic theory, $\mathcal
{B}$ is the budget set
associated with an agent with unit endowment, when prices of bundles in
${\mathbb{L}^0_+}$ are given in a linear way by $\mu$: the price of
$f \in{\mathbb{L}^0_+}$
is simply $\int_\Omega f \,d\mu$.

The concept of a full simplex naturally incorporates num\'{e}raire-invariance.
If $\mathcal{B}$ is a full simplex in ${\mathbb{L}^0_+}$ and $f \in
{\mathbb{L}^0_{++}}$, then $(1 / f)
\mathcal{B}
$ is also a full simplex in ${\mathbb{L}^0_+}$. In fact, and in view
of the
characterization given in statement (3) of Theorem
\ref{thm:geomcharofLone}, starting from a full simplex $\mathcal{B}$ in
${\mathbb{L}^0_+}$,
the class of
sets of the form $(1 / f) \mathcal{B}$, where $f$ ranges in ${\mathbb
{L}^0_{++}}$, coincides
with the class of \textit{all} the full simplices in ${\mathbb
{L}^0_+}$. Therefore,
the class of full simplices in ${\mathbb{L}^0_+}$ has the same
cardinality as
${\mathbb{L}^0_{++}}$.

To further get a feeling for the ``fullness'' of full simplices, we
mention the following result. Apart from its independent interest, it
will be crucial in proving the axiomatic characterization of
num\'{e}raire-invariant choices given in Theorem~\ref{thm:converse}.
\begin{prop} \label{prop:fullnessoffullsimplices}
Let $\mathcal{B}$ be a full simplex in ${\mathbb{L}^0_+}$ and
$\mathcal{C}$ be a convex subset of
${\mathbb{L}^0_+}$ such that $\mathcal{B}\subseteq\mathcal{C}$ and
$\mathcal{B}^{\max}\cap\mathcal{C}^{\max}\cap{\mathbb{L}^0_{++}}
\neq\varnothing$. Then, actually, $\mathcal{B}= \mathcal{C}$.
\end{prop}
\begin{pf}
Pick $h \in\mathcal{B}^{\max}\cap\mathcal{C}^{\max}\cap{\mathbb
{L}^0_{++}}$. Replacing $\mathcal{B}$ and $\mathcal{C}$ with
$(1 / h) \mathcal{B}$ and $(1 / h) \mathcal{C}$, respectively, we may
assume that $\mathcal{C}
\subseteq{\mathbb{L}^0_+}$ is convex, $\mathcal{B}\subseteq\mathcal
{C}$, $\mathcal{B}$ is a full simplex in
${\mathbb{L}^0_+}$ and $1 \in\mathcal{B}^{\max}\cap\mathcal
{C}^{\max}$. Furthermore, we can assume that
$\mathcal{C}
$ is solid, replacing it if necessary with $\{f \in{\mathbb
{L}^0_+}\mid f \leq g \mbox{ for some } g \in\mathcal{C} \}$, since
all the above properties
will still hold. By Theorem \ref{thm:geomcharofLone}, there exists
a $\sigma$-finite measure $\mu\sim\Pi$ such that $\mathcal{B}= \{f
\in {\mathbb{L}^0_+}\mid\int_\Omega f \,d\mu\leq1 \}$. As $1 \in
\mathcal{B}^{\max}$, it is easy
to see that $\mu$ has to actually be a probability, which we then
denote by $\mathbb{P}$; that is, $\mathcal{B}= \{f \in{\mathbb
{L}^0_+}\mid\mathbb{E}_\mathbb{P}[f] \leq 1 \}$. All the previous
assumptions and notation will be in force in the
sequel. We have to show that $\mathcal{B}= \mathcal{C}$.

For $n \in\mathbb N$, define a convexly compact set $E^n$ as the
closure of $\mathcal{C}\cap\{ f \in{\mathbb{L}^0_+}\mid f \leq n \}
$. With $\preccurlyeq_\mathbb{P}$
defined via (\ref{eq:preferenceviaQ}), for each $n \in\mathbb N$ let
$h^n \in E^n$ satisfy $f \preccurlyeq_\mathbb{P}h^n$ for all $f \in
E^n$. If $h^n =
1$ for all $n \in\mathbb N$, then $\mathbb{E}_\mathbb{P}[f] \leq1$
for all $\mathcal{C}
\cap
{\mathbb{L}^\infty_+}$ and, by Fatou's lemma and the solidity of
$\mathcal{C}$, $\mathbb{E}_\mathbb{P}[f]
\leq
1$ for all $\mathcal{C}$; therefore, $\mathcal{C}\subseteq\mathcal
{B}$ and there is nothing left
to prove. By way of contradiction, assume that $\mathbb{P}[h^n = 1] <
1$ for
some $n \in\mathbb N$; then, a fortiori, $n \geq2$. Note then that
$\mathbb{E}_\mathbb{P}[h^n] > 1$, that is, $h^n \notin\mathcal{B}$,
which follows from the
facts that $\mathbb{E}_\mathbb{P}[1 / h^n] \leq1$ (since $1 \in
E^n$) and $\mathbb{P}[h^n
= 1] < 1$. From now onward, fix $n \in\mathbb N$ with $n \geq2$ such
that $h^n$ has the previous property, and we drop the superscript
``$n$'' from everywhere for typographical convenience. Let also $D :=
\mathcal{B}\cap\{ f \in{\mathbb{L}^0_+}\mid f \leq n \}$. Remember
throughout that the
elements of $D$ and $E$ are included in the $\mathbb{L}^\infty$-ball
of radius
$n$, that $D \subseteq E$, and that $h \in E \setminus D$.

Let $\pi$ be the $\mathbb{L}^2(\mathbb{P})$-projection of $h$ on
$D$---observe
that this is well defined since all elements of $E$ (and therefore also
of $D \subseteq E$) belong to $\mathbb{L}^\infty\subseteq\mathbb
{L}^2(\mathbb{P})$ and
$D$ is convex and $\mathbb{L}^2(\mathbb{P})$-closed. Also, let $\nu
:= h - \pi$.
Since $h \notin D$, $\mathbb{P}[\nu= 0 ] < 1$. Define $\pi' := \pi
\mathbb{I}
_{\{ \nu\geq0\}} + h \mathbb{I}_{\{ \nu< 0\}}$. Since $h < \pi$ on
$\{
\nu< 0\}$, we have $\pi' \leq\pi$, which implies in particular that
$\pi' \in D$. Also, since $\{\pi' < \pi \} = \{\nu< 0 \}$,
$\mathbb{P}[
\nu< 0 ] > 0$ would imply $\mathbb{E}_\mathbb{P}[|\pi' - h|^2 ] =
\mathbb{E}_\mathbb{P}[|\pi- h|^2 \mathbb{I}_{\{\nu\geq0 \}} ] <
\mathbb{E}_\mathbb{P}[|\pi- h|^2 ]$,
which contradicts the fact that $\pi$ is the $\mathbb{L}^2(\mathbb
{P})$-projection
of $h$ on $D$. Therefore, $\nu\in{\mathbb{L}^\infty_+}$.

Define
\[
\delta:= \min\biggl\{\frac{\mathbb{E}_\mathbb{P}[h] - 1}{\mathbb
{E}_\mathbb{P}[\nu]}, 1 \biggr\} \in(0,
1]
\]
as well as
\[
\zeta:= 1 + \frac{1}{n} - \frac{1}{n}
(h - \delta\nu) = 1 + \frac{1}{n} - \frac{1}{n} \bigl( \pi+ (1
-\delta) \nu\bigr).
\]
The above definition of $\delta$ ensures that $\mathbb{E}_\mathbb
{P}[\zeta] \leq1$.
Also, $0 \leq\pi= h - \nu\leq h - \delta\nu\leq h \leq n$, which
implies that $\mathbb{P}[1/n \leq\zeta\leq1 + 1/n] = 1$, and, therefore,
that $\zeta\in D$, since $n \geq2$. If $\zeta\in E$, then also $1 +
\delta\nu/ (n + 1) = ((n / (n + 1) ) \zeta+ (1 / (n+1) ) h ) \in E$,
which is impossible in view of $1 \in E^{\max}$ ($1
\in E \subseteq\mathcal{C}$ and $1 \in\mathcal{C}^{\max}$). We
obtain that $\zeta\in D
\setminus E$, which is a contradiction to the fact that $D \subseteq
E$. The last contradiction implies that $\mathbb{P}[h \neq1] > 0$ is
impossible, which concludes the proof.
\end{pf}

\subsection{\texorpdfstring{Axiomatic characterization of num\'{e}raire-invariant
choices}{Axiomatic characterization of numeraire-invariant choices}}

\subsubsection{The characterization result} We are ready to give the
main result of this section.
\begin{theorem} \label{thm:converse}
Let $\preccurlyeq$ be a binary relation on ${\mathbb{L}^0_+}$ that
satisfies the
following properties:
\begin{enumerate}[(A4)]
\item[(A1)] $f \preccurlyeq g$ holds if and only if $\{f > 0 \}
\subseteq
\{g > 0 \}$ and $(f / g) \mathbb{I}_{\{g > 0 \}} + \mathbb{I}_{\{g =
0 \}}
\preccurlyeq1$.
\item[(A2)] If $f \leq1$, then $f \preccurlyeq1$. Furthermore, if $f
\leq
1$ and $\{f < 1 \}$ is not $\Pi$-null, then $f \prec1$.
\item[(A3)] The lower-contour set $\{ f \in{\mathbb{L}^0_+}\mid f
\preccurlyeq1 \}$
is convex.
\item[(A4)] For some full simplex $\mathcal{B}$ of ${\mathbb
{L}^0_+}$, there exists $\widehat{f}
\in
\mathcal{B}$ such that $f \preccurlyeq\widehat{f}$ holds for all $f
\in\mathcal{B}$.\vadjust{\goodbreak}
\end{enumerate}
Then, there exists a unique $\mathbb{P}\in\Pi$ that \textit{generates}
$\preccurlyeq$, in the sense that $\preccurlyeq$ is exactly the relation
$\preccurlyeq_\mathbb{P}
$ of (\ref{eq:preferenceviaQ}).
\end{theorem}
\begin{pf}
For \textit{any} $\mathbb{Q}\in\Pi$, let $\mathcal{B}^1_\mathbb
{Q}:= \{f \in{\mathbb{L}^0_+} \mid\mathbb{E}_\mathbb{Q}[f] \leq1
\}$. Also let $\mathcal{C}^{1}_{\preccurlyeq}:= \{f \in{\mathbb
{L}^0_+} \mid f \preccurlyeq1 \}$. By the num\'{e}raire-invariance
axiom (A1), proving Theorem
\ref{thm:converse} amounts to finding $\mathbb{P}\in\Pi$ such that
$\mathcal{B}^1_\mathbb{P}= \mathcal{C}^{1}_{\preccurlyeq}$.

A combination of (A1) and (A4) imply that for \textit{any} full simplex
$\mathcal{B}$ of ${\mathbb{L}^0_+}$, there exists $\widehat{f}\in
\mathcal{B}$ such that $f \preccurlyeq\widehat{f}$
holds for all $f \in\mathcal{B}$. Fix $\mathbb{Q}\in\Pi$. By
Theorem \ref{thm:geomcharofLone}, $\mathcal{B}^1_\mathbb{Q}$ is a full simplex in
${\mathbb{L}^0_+}$; therefore,
there exists $g \in\mathcal{B}^1_\mathbb{Q}$ such that $f
\preccurlyeq g$ holds for all $f
\in\mathcal{B}^1_\mathbb{Q}$. We claim that $g \in{\mathbb
{L}^0_{++}}$, as well as $\mathbb{E}_\mathbb{Q}[g] = 1$.
Indeed, $g \in{\mathbb{L}^0_{++}}$ follows from the fact $\mathcal
{B}^1_\mathbb{Q}\ni1 \preccurlyeq g$,
since (A1) implies that in this case $\Omega= \{1 > 0 \} \subseteq\{
g > 0 \}$. Also, if $\mathbb{E}_\mathbb{Q}[g] < 1$, then $h :=
(\mathbb{E}_\mathbb{Q}[g])^{-1} g
\in\mathcal{B}^1_\mathbb{Q}$ with $\mathbb{P}[g < h] = 1$, which
means that $g \prec h$ by
(A2) and contradicts the fact that $h \preccurlyeq g$ for $h \in
\mathcal{B}^1_\mathbb{Q}$.

Define $\mathbb{P}\in\Pi$ via $\mathbb{P}[A] := \mathbb{E}_\mathbb
{Q}[g \mathbb{I}_A]$ for all
$A \in\mathcal{F}$. Observe that $f \in\mathcal{B}^1_\mathbb{P}$
if and only if $(f g) \in
\mathcal{B}^1_\mathbb{Q}$, and in that case we have $f g \preccurlyeq
g$, or $f \preccurlyeq1$ in
view of axiom (A1). In other words, $\mathcal{B}^1_\mathbb{P}\subseteq
\mathcal{C}^{1}_{\preccurlyeq}$. Since
$\mathcal{C}^{1}_{\preccurlyeq}$ is convex by (A3), and $1 \in
(\mathcal{B}^1_\mathbb{P})^{\max} \cap
(\mathcal{C}^{1}_{\preccurlyeq})^{\max} \cap{\mathbb{L}^0_{++}}$,
where $1 \in(\mathcal{C}^{1}_{\preccurlyeq})^{\max}$
follows from (A2), an application of Proposition \ref
{prop:fullnessoffullsimplices} gives $\mathcal{B}^1_\mathbb{P}=
\mathcal{C}^{1}_{\preccurlyeq}$.

We finally discuss the uniqueness of the representative $\mathbb{P}\in
\Pi
$. If $\mathbb{P}' \in\Pi$ also generates $\preccurlyeq$, then
$\mathcal{B}^1_\mathbb{P}=
\mathcal{C}^{1}_{\preccurlyeq}= \mathcal{B}^1_{\mathbb{P}'}$ should
hold, which implies that $\mathbb{P}=
\mathbb{P}'$, and completes the proof.
\end{pf}

A comparison with the statement of Theorem \ref{thm:staticproperties}
is in order. Axioms (A1) and (A2) of Theorem \ref{thm:converse} are
really the same as statements (1) and (2) of
Theorem~\ref{thm:staticproperties}---it is enough to deal with the case $g = 1$
in axiom
(A2) of Theorem \ref{thm:converse} because of the num\'{e}raire-invariance
axiom (A1). The first surprise comes from the simplicity of axiom (A3)
of Theorem \ref{thm:converse}, where we \textit{only} require convexity
of the lower contour set. This should be compared to the very rich
structure that is given in statement (3) of Theorem
\ref{thm:staticproperties} for both the lower-contour and upper-contour
sets. The num\'{e}raire-invariance axiom (A1) is strong enough so that
\textit{no} closedness or
even risk-aversion axiom is needed. Also, axiom (A4) of
Theorem~\ref{thm:converse} is significantly weaker than statement (4) of
Theorem \ref{thm:staticproperties}, as it only asks that an optimal choice
exists for \textit{some} full simplex of ${\mathbb{L}^0_+}$, and not
for \textit{all}
convexly compact subsets of ${\mathbb{L}^0_+}$. Although, in view of
(A1), (A4)
actually implies that an optimal choice exists for \textit{all} full
simplices of ${\mathbb{L}^0_+}$; this class is still much smaller than
the class of
all convexly compact sets.

\subsubsection{Subjective probability and risk aversion}

The probability $\mathbb{P}\in\Pi$ that generates the relation
$\preccurlyeq$
satisfying the axioms of Theorem \ref{thm:converse} should be thought
as the subjective probability of the agent whose choices are
represented by $\preccurlyeq$, as it corresponds to the idea of
``agent risk
aversion.'' If the agent's subjective probability is $\mathbb{Q}\in
\Pi$,
risk aversion would translate into $f \preccurlyeq\mathbb{E}_\mathbb
{Q}[f]$ holding for
all $f \in{\mathbb{L}^\infty_+}$. Let $\mathbb{P}\in\Pi$ generate
$\preccurlyeq$. Then,
$\mathbb{P}
[A] / \mathbb{Q}[A] = \mathbb{E}_\mathbb{P}[(1 / \mathbb{E}_\mathbb
{Q}[\mathbb{I}_A]) \mathbb{I}_A] \leq1$,
that is, $\mathbb{P}[A] \leq\mathbb{Q}[A]$, holds for all nonnull
$A \in\mathcal{F}$.
Therefore, $\mathbb{Q}= \mathbb{P}$.

\subsubsection{Choice rules}
A more behavioral-based alternative to modeling preferences via binary
relations is to model the \textit{choice rules} of an agent; for a quick
introduction and the material we shall need here, see Chapter 1 of
\cite{mascolellwhinstongreen}. For all $\mathcal{C}\subseteq{\mathbb
{L}^0_+}$, define $\varepsilon_\preccurlyeq
(\mathcal{C}) := \{g \in\mathcal{C}\mid f \preccurlyeq g, \mbox{
for all } f \in\mathcal{C} \}$. This way we get a \textit{choice
function} $\varepsilon= \varepsilon_\preccurlyeq$. Forgetting
that $\varepsilon$ came from $\preccurlyeq$, we can define the \textit{revealed}
preference $\preccurlyeq_\varepsilon$ from $\varepsilon$ as follows:
$f \preccurlyeq_\varepsilon g$ if
and only if there exists $\mathcal{C}\subseteq{\mathbb{L}^0_+}$ such
that $f \in\mathcal{C}$ and
$g \in\varepsilon(\mathcal{C})$. Then, it can be shown that
$\preccurlyeq_\varepsilon$ coincides
with $\preccurlyeq$ on ${\mathbb{L}^0_+}$. Furthermore, the axioms of
Theorem \ref{thm:converse} can be expressed directly in terms of the choice rule
$\varepsilon$;
therefore, this can be viewed as as the starting point of
axiomatization, which will then induce the preference structure
$\preccurlyeq$.

\subsection{Extending the preference structure}

As noted in Section \ref{subsubsec:nottransitive}, one of the
``drawbacks'' of a preference relation that satisfies the axioms of
Theorem \ref{thm:converse} is that it fails to be transitive. We shall
extend $\preccurlyeq$ to a preference relation $\trianglelefteq$ that
is transitive
and satisfies some extremely weak continuity properties. To avoid
unnecessary technicalities, we shall work on ${\mathbb{L}^0_{++}}$. As
it will turn
out, $\trianglelefteq$ \textit{almost} has a numerical representation
given by
expected logarithmic utility under the probability $\mathbb{P}\in\Pi
$ that
generates $\preccurlyeq$. We shall discuss the previous use of the word
``almost'' after stating and proving Theorem \ref{thm:preferences} below.

As with any preference relation, $f \vartriangleleft g$ will mean that
$f \trianglelefteq
g$ holds, whereas $g \trianglelefteq f$ fails to hold. Also, for $x \in
\mathbb R
_{++}$, we set $\log_+(x) = \max\{\log(x), 0 \}$.
\begin{theorem} \label{thm:preferences}
Let $\preccurlyeq$ denote a binary relation on ${\mathbb{L}^0_+}$
satisfying the axioms
of Theorem \ref{thm:converse}. Then, there exists a (not necessarily
unique) binary relation $\trianglelefteq$ on ${\mathbb{L}^0_{++}}$
such that:
\begin{enumerate}[(3)]
\item[(1)] If $f \in{\mathbb{L}^0_{++}}$ and $g \in{\mathbb{L}^0_{++}}$,
$f \trianglelefteq g$ holds if and
only if $(f / g) \trianglelefteq1$.
\item[(2)] For $f \in{\mathbb{L}^0_{++}}$, $f \prec1$ implies $f
\vartriangleleft1$.
\item[(3)] $\trianglelefteq$ is transitive.
\item[(4)] For $f \in{\mathbb{L}^0_{++}}$, $f \trianglelefteq1$ is
implied by either of the
conditions below:
\begin{enumerate}[(a)]
\item[(a)] $a f \trianglelefteq1$ holds for all $a \in(0,1)$.
\item[(b)] $f \geq\epsilon$ for some $\epsilon\in\mathbb R_{++}$, and $f
\wedge n \trianglelefteq1$ holds for all $n \in\mathbb N$.
\end{enumerate}
\end{enumerate}
In this case, and with $\mathbb{P}\in\Pi$ generating $\preccurlyeq
$, the
following holds: for any $f \in{\mathbb{L}^0_{++}}$ and $g \in
{\mathbb{L}^0_{++}}$ with
$\mathbb{E}_\mathbb{P}
[\log_+ (f / g ) ] < \infty$, we have
%
\begin{equation} \label{eq:preferencerepresentation}
f \trianglelefteq g \quad\Longleftrightarrow\quad\mathbb{E}_\mathbb{P}\biggl[\log
\biggl(\frac {f}{g} \biggr) \biggr] \leq0.
\end{equation}
As a corollary, the restriction of any binary relation $\trianglelefteq$
satisfying (1), (2), (3) and (4) above on ${\mathcal{L}}_\mathbb{P}:=
\{f \in {\mathbb{L}^0_{++}}\mid\mathbb{E}_\mathbb{P}[|\log f| ] <
\infty \}$ is uniquely defined via
the numerical representation
\[
\mbox{for } f \in{\mathcal{L}}_\mathbb{P}\mbox{ and } g \in
{\mathcal{L}}_\mathbb{P}\qquad f
\trianglelefteq g \quad\Longleftrightarrow\quad\mathbb{E}_\mathbb{P}[\log(f)
] \leq\mathbb{E}_\mathbb{P}[\log(g) ].
\]
In particular, $\trianglelefteq$ is complete on ${\mathcal
{L}}_\mathbb{P}$.
\end{theorem}
\begin{pf}
We shall first establish the existence of a binary relation
$\trianglelefteq$ on
${\mathbb{L}^0_{++}}$ that satisfies the requirements (1), (2), (3)
and (4) of
Theorem \ref{thm:preferences}. We use the following definition: for $f
\in{\mathbb{L}^0_{++}}$ and $g \in{\mathbb{L}^0_{++}}$, we
\textit{set $f \trianglelefteq g$ if and only if
$\mathbb{E}_\mathbb{P}[\log_+(f / g)] < \infty$ and $\mathbb
{E}_\mathbb{P}[\log(f / g)] \leq0$
hold}. The num\'{e}raire-invariance property (1) and the transitivity property
(3) are straightforward.
For property (2), note that if $f \prec1$, that is, $\mathbb
{E}_\mathbb{P}[f] < 1$,
for $f \in{\mathbb{L}^0_{++}}$, Jensen's inequality implies that
$\mathbb{E}_\mathbb{P}[\log
(f)] <
0 = \mathbb{E}_\mathbb{P}[\log(1)]$, that is, $f \vartriangleleft1$.
Finally, property (4a) is trivial to check, while property (4b) follows
from the monotone convergence theorem.

Conversely, consider \textit{any} binary relation that satisfies all the
requirements of Theorem \ref{thm:preferences}. First of all, we claim
that $f \trianglelefteq1$ and $g \trianglelefteq1$ imply that $f g
\trianglelefteq1$. Indeed,
$g \trianglelefteq1$ is equivalent to $1 \trianglelefteq1 / g$ by the
num\'{e}raire-invariance
property (1), and then the transitivity property (3) gives $f
\trianglelefteq1
/ g$. The num\'{e}raire-invariance property (1) applied once again
gives $f g
\trianglelefteq1$.

We now show that $f \vartriangleleft g$ and $g \trianglelefteq h$ imply
$f \vartriangleleft h$. We
already know that $f \trianglelefteq h$ from the transitivity property
(3). If
$h \trianglelefteq f$, then $h / f \trianglelefteq1$ and $g / h
\trianglelefteq1$ would imply
$(h/f)(g/h) \trianglelefteq1$, or $g/f \trianglelefteq1$, or again
equivalently that $g
\trianglelefteq f$, which is false. Therefore, $f \vartriangleleft h$.

Pick $f \in{\mathbb{L}^0_{++}}$ such that $f \leq M$ for some $M \in
\mathbb R_{+}$ and
$\mathbb{E}_\mathbb{P}[\log(f)] < 0$. Define $\ell^n := n (f^{1/n}
-1)$ for all $n
\in\mathbb N$. Then, $\downarrow\lim_{n \to\infty}\ell^n = \log
(f)$ and $\ell^n
\leq\ell^1 \leq M - 1$ for all $n \in\mathbb N$. Therefore, the
monotone convergence theorem gives that $\mathbb{E}_\mathbb{P}[\ell
^n] < 0$ for some
large enough $n \in\mathbb N$. This means that $\mathbb{E}_\mathbb
{P}[f^{1/n}] \leq
1$. As $f \neq1$ (which follows from $\mathbb{E}_\mathbb{P}[\log
(f)] < 0$), we have
$f^{1 / n} \prec1$, that is, $f^{1/n} \vartriangleleft1$ by the extension
property (2), and therefore, $f \vartriangleleft1$ by the results of the
preceding paragraphs.

Pick $f \in{\mathbb{L}^0_{++}}$ with $\mathbb{E}_\mathbb{P}[\log
(f)] < 0$. Choose $\epsilon\in
\mathbb R_{++}$ such that $\mathbb{E}_\mathbb{P}[\log(f + \epsilon
)] < 0$. Then,
$\mathbb{E}_\mathbb{P}
[\log( (f + \epsilon) \wedge M)] < 0$ holds for all $M \in\mathbb R_{++}$;
therefore, $(f + \epsilon) \wedge M \trianglelefteq1$ holds for all
$M \in
\mathbb R
_{++}$ by the result of the preceding paragraph. Since $f + \epsilon
\geq\epsilon$, the weak continuity property (4b) gives $(f + \epsilon)
\trianglelefteq1$. Finally, since $f \prec f+\epsilon$, we have $f
\vartriangleleft f +
\epsilon$ by the extension property (2), which combined with $(f +
\epsilon) \trianglelefteq1$ gives $f \vartriangleleft1$.

Up to now, we have shown that $f \in{\mathbb{L}^0_{++}}$ with
$\mathbb{E}_\mathbb{P}[\log(f)] < 0$
implies $f \vartriangleleft1$. Pick $f \in{\mathbb{L}^0_{++}}$ with
$\mathbb{E}_\mathbb{P}[\log(f)] \leq0$.
Then, for all $a \in(0,1)$ we have $\mathbb{E}_\mathbb{P}[\log(a
f)] < 0$;
therefore, $a f \trianglelefteq1$. The continuity property (4a) gives
$f \trianglelefteq
1$. Therefore, $f \in{\mathbb{L}^0_{++}}$ with $\mathbb{E}_\mathbb
{P}[\log(f)] \leq0$ implies $f
\trianglelefteq1$.

Finally, pick $f \in{\mathbb{L}^0_{++}}$ with $\mathbb{E}_\mathbb
{P}[\log_+(f)] < \infty$, and
assume that $f \trianglelefteq1$. Then, we claim that we must have
$\mathbb{E}_\mathbb{P}
[\log
(f)] \leq0$. Suppose on the contrary that $\mathbb{E}_\mathbb
{P}[\log(f)] > 0$; this
would imply that $1 \vartriangleleft f$, which is impossible.
Therefore, for $f
\in{\mathbb{L}^0_{++}}$ with $\mathbb{E}_\mathbb{P}[\log_+(f)] <
\infty$ we have that $f
\trianglelefteq1$
if and only if $\mathbb{E}_\mathbb{P}[\log(f)] \leq0$, which is
exactly what we
needed to show.
\end{pf}

The special relation $\trianglelefteq$ constructed in the first
paragraph of the
proof of Theorem~\ref{thm:preferences} is the \textit{minimal} way to
construct a binary relation on ${\mathbb{L}^0_{++}}$ that satisfies
the requirements
(1), (2), (3) and (4) of Theorem \ref{thm:preferences}; any other such
relation has to be an extension of the one described there. Observe
that if $\mathbb{L}^0$ is finite-dimensional, ${\mathcal{L}}_\mathbb
{P}= {\mathbb{L}^0_{++}}$ and therefore
in this case we obtain the uniqueness of $\trianglelefteq$ that
satisfies the
requirements (1), (2), (3) and (4) of Theorem \ref{thm:preferences}.

Theorem \ref{thm:preferences} remains silent on how to define the
relation between $f \in{\mathbb{L}^0_{++}}$ and $g \in{\mathbb
{L}^0_{++}}$ when both $\mathbb{E}_\mathbb{P}
[\log
_+(f/g)] = \infty$ and $\mathbb{E}_\mathbb{P}[\log_+(g/f)] = \infty
$ hold. (When
$\mathbb{L}^0
$ is infinite-dimensional, one can always find pairs like this.)
Note that, for $f \in{\mathbb{L}^0_{++}}$ such that $\mathbb
{E}_\mathbb{P}[\log_+(f)] < \infty$,
$\mathbb{E}_\mathbb{P}[\log_+(1/f)] = \infty$ implies $f
\trianglelefteq1$ by (\ref{eq:preferencerepresentation}).
One would be tempted to define $f \trianglelefteq1$ whenever $\mathbb
{E}_\mathbb{P}[\log
_+(1/f)] = \infty$, claiming that there is too much ``downside risk''
in $f$. However, with this understanding, if $f \in{\mathbb
{L}^0_{++}}$ is such that
$\mathbb{E}_\mathbb{P}[\log_+(f)] = \mathbb{E}_\mathbb{P}[\log
_+(1 / f)] = \infty$, we would
get $f
\trianglelefteq1$ and $1/f \trianglelefteq1$, or equivalently that $f
\trianglelefteq1$ and $1
\trianglelefteq f$, which would make all $f \in{\mathbb{L}^0_{++}}$
such that $\mathbb{E}_\mathbb{P}[\log
_+(f)] = \mathbb{E}_\mathbb{P}[\log_+(1 / f)] = \infty$ belong to
the same
equivalence class. This is impossible: if $f \in{\mathbb{L}^0_{++}}$
is such that
$\mathbb{E}_\mathbb{P}[\log_+(f)] = \mathbb{E}_\mathbb{P}[\log
_+(1 / f)] = \infty$, then $2 f$ has
the same property, but $f \vartriangleleft2 f$. We may simply opt to
leave the
relation of $f$ and $g$ when $\mathbb{E}_\mathbb{P}[\log_+(f / g)] =
\mathbb{E}_\mathbb{P}[\log_+(g
/ f)] = \infty$ undefined, implicitly claiming that they are too risky
relatively to each other to be compared. It remains an \textit{open
question} whether one can extend $\trianglelefteq$ to make it complete on
${\mathbb{L}^0_{++}}
$, still having the properties of Theorem \ref{thm:preferences}
holding, when $\mathbb{L}^0$ is infinite-dimensional.

\section{\texorpdfstring{Num\'{e}raire-invariant preferences in a dynamic
environment}{Numeraire-invariant preferences in a dynamic environment}}
\label{sec:dynamic}

\subsection{Notation and terminology}
All stochastic processes in the sequel are defined on a filtered
probability space $(\Omega, (\mathcal{F}_t)_{t \in\mathbb R_+},
\mathbb{P})$. Here, the probability $\mathbb{P}$ on $(\Omega,
\mathcal{F}_\infty)$, where $\mathcal{F}_\infty:= \bigvee_{t \in
\mathbb R_+} \mathcal{F}_t$ will
be fixed and we shall be using ``$\mathbb{E}$'' for the expectation of
$\mathcal{F}
_\infty$-measurable random variables under $\mathbb{P}$. The filtration
$(\mathcal{F}
_t)_{t \in\mathbb R_+}$ is assumed to be right-continuous and
$\mathcal{F}_0$ is
assumed $\mathbb{P}$-trivial. The optional $\sigma$-algebra on
$\Omega\times\mathbb R_+$ is
denoted by $\mathcal{O}$. A set $A \in\mathcal{O}$ is called \textit{evanescent} if
the random set $\Omega\ni\omega\mapsto\{t \in\mathbb R_+ \mid(t,
\omega) \in A \}$ is $\mathbb{P}$-a.s. empty; an optional process $V$ is
evanescent if $\{V \neq0 \} \in\mathcal{O}$ is an evanescent set.
For $A
\in\mathcal{O}$ and $t \in\mathbb R_+$, we set $A_t := \{\omega\in
\Omega \mid(\omega, t) \in A \} \in\mathcal{F}_t$.

For a c\`{a}dl\`{a}g process $X$ we define the process $X_-= (X_{t-})_{t
\in\mathbb R_+}$ by $X_{0-} = 0$, and $X_t$ being the left-limit of
$X$ at
$t \in\mathbb R_{++}$. Also, we let $\Delta X := X - X_-$. Every
predictable process $H$ is supposed to satisfy $H_0 = 0$. Whenever $H$
and $X$ are $d$-dimensional processes such that $X$ is a semimartingale
to be used as an integrator, and $H$ can be used as integrand with
respect to $X$, we denote by $\int_{[0 ,\cdot]} \langle H_t , dX_t
\rangle$
the integral process, where ``$\langle\cdot , \cdot \rangle$'' is
used to
(sometimes, formally) denote the usual inner product in $\mathbb R^d$. We
assume vector stochastic integration (see, e.g.,
\cite{MR1943877}). Note that $\int_{\{0 \}} \langle H_t , dX_t \rangle =
\langle H_0 , \Delta X_0 \rangle = \langle H_0 , X_0 \rangle$;
therefore, if $H$ is
predictable, $\int_{\{0 \}} \langle H_t , dX_t \rangle = 0$. We also define
$\int_{(0 ,\cdot]} \langle H_t , dX_t \rangle := \int_{[0 ,\cdot]}
\langle H_t , dX_t \rangle - \int_{\{0 \}} \langle H_t , dX_t
\rangle = \int_{[0
,\cdot
]} \langle H_t , dX_t \rangle - \langle H_0 , X_0 \rangle$.

\subsection{A canonical representation of unit-mass optional measures}
The natural space to define ``subjective probabilities'' of agents in
the dynamic case is $(\Omega\times\mathbb R_+, \mathcal{O})$. We
begin with a result regarding
the structure of nonnegative measures on $(\Omega\times\mathbb R_+,
\mathcal{O})$ with unit
total mass.
\begin{theorem} \label{thm:doleanssharpened}
On $(\Omega\times\mathbb R_+, \mathcal{O})$, consider a measure $p$
such that $p[\Omega\times\mathbb R_+] = 1$
and $p[A] = 0$ for every evanescent set $A \in\mathcal{O}$. Then, there
exists a pair of processes $(L, K)$ such that:
\begin{enumerate}[(1)]
\item[(1)] $L$ is a nonnegative local martingale with $L_0 = 1$.
\item[(2)] $K$ is adapted, right-continuous, nondecreasing, and $0 \leq K
\leq1$.
\item[(3)] $\int_{\Omega\times\mathbb R_+} V \,dp = \mathbb{E}[\int
_{\mathbb R_+} V_t L_t \,dK_t ]$ holds for all nonnegative optional
process $V$.
\item[(4)] $L = \int_{[0 ,\cdot]}\mathbb{I}_{\{K_{t-} < 1 \}} \,dL_t$ and
$K =
\int_{[0 ,\cdot]}\mathbb{I}_{\{L_{t} > 0 \}} \,dK_t$.
\end{enumerate}
Furthermore, $\{L_\infty> 0 \} \subseteq\{K_\infty= 1 \}$ holds.

A pair $(L, K)$ that satisfies the above requirements is essentially
unique, in the following sense: if $(K', L')$ is another pair that
satisfies the above requirements, then $K = K'$ up to evanescence,
while $L_t = L'_t$ for all $t \in\mathbb R_+$ holds on $\{K_\infty> 0
\}$.
\end{theorem}
%
%
\begin{defn}
For a measure $p$ on $(\Omega\times\mathbb R_+, \mathcal{O})$ with
$p[\Omega\times\mathbb R_+] = 1$ and $p[A]
= 0$ holding for every evanescent set $A \in\mathcal{O}$, a pair of processes
$(L, K)$ that satisfies requirements (1), (2), (3) and (4) of
Theorem \ref{thm:doleanssharpened} will be called a \textit{canonical
representation pair for $p$}.
\end{defn}
\begin{rem} \label{rem:subjectiveviews1}
Let $p \in\Pi$ with canonical representation pair $(L, K)$, and
suppose that $L$ is the density process of a probability $\mathbb{Q}$ with
respect to $\mathbb{P}$; for this, it is necessary that $L$ is a martingale
and sufficient that $L$ is a uniformly integrable martingale. For all
$t \in\mathbb R_+$ and $A \in\mathcal{F}_t$, $\mathbb{Q}[A] =
\mathbb{E}[L_t \mathbb{I}_A]$,
that is, $\mathbb{Q}$ is locally absolutely continuous with respect to
$\mathbb{P}$. Furthermore, using integration-by-parts and a standard
localization argument, it is straightforward to check that $\int
_{\Omega\times\mathbb R_+} V \,dp = \mathbb{E}_\mathbb{Q}[\int
_{\mathbb R_+} V_t \,dK_t ]$ holds
for all nonnegative optional process~$V$. Since $p [\Omega\times
\mathbb R
_+] = 1$ and $\mathbb{Q}[K_\infty\leq1] = 1$ hold, it must be the case
that $\mathbb{Q}[K_\infty= 1] = 1$.

As it turns out, however, the above special case is not exhaustive. It
may happen that $L$ is a \textit{strict} local martingale in the sense of
\cite{MR1478722}, which precludes it from being a density process of
some probability $\mathbb{Q}$ with respect to $\mathbb{P}$.
(Nevertheless, at
least in the case of finite time-horizon, one is able to interpret $L$
as the density process of a \textit{finitely} additive probability with
respect to $\mathbb{P}$, that is only locally countably additive (for more
information, see \cite{MR2114989}).) It might also happen that $\{
K_\infty< 1 \}$ is not $\mathbb{P}$-null; actually, it can even
happen that
$\mathbb{P}[K_\infty< 1] = 1$. The previous are illustrated in Example
\ref{exa:coolstuff} later on in the text.
\end{rem}

\subsection[Existence of a canonical representation pair in Theorem 2.1]{Existence
of a canonical representation pair in Theorem
\protect\ref{thm:doleanssharpened}}
\label{subsec:proofofthm:doleanssharpened}

Dol\'{e}ans's representation of optional measures (see, e.g., Section
VI.20 of \cite{MR1780932}) implies the existence of an adapted,
right-continuous, nonnegative and nondecreasing process $H$ such that
$\int_{\Omega\times\mathbb R_+} V \,dp = \mathbb{E}[\int_{\mathbb
R_+} V_t \,dH_t ]$ for
all nonnegative optional processes $V$. We shall establish below that
any adapted, right-continuous, nonnegative and nondecreasing process
$H$ with $\mathbb{E}[H_\infty] = 1$ can be decomposed as $H = \int_{[0,
\cdot]} L_t \,dK_t$ for a pair $(L, K)$ satisfying (1), (2), and (4)
of Theorem \ref{thm:doleanssharpened}. The question of essential
uniqueness of the pair $(L, K)$ satisfying properties (1), (2), (3) and
(4) of Theorem \ref{thm:doleanssharpened} will be tackled in Section
\ref{subsec:uniquenessofcanreppair}.

Consider the nonnegative c\`{a}dl\`{a}g martingale $M$ that satisfies
$M_t =
\mathbb{E}[H_\infty\mid\mathcal{F}_t]$ for all $t \in\mathbb R_+$.
Then, define the
supermartingale $Z := M - H$; $Z$ is nonnegative since $Z_t = \mathbb{E}
[H_\infty- H_t \mid\mathcal{F}_t]$ holds for all $t \in\mathbb
R_+$. The
expected total mass of $H$ over $\mathbb R_+$ is $M_0 = \mathbb
{E}[H_\infty] =
1$. If $\mathbb{P}[H_\infty> 1] = 0$, in which case $\mathbb
{P}[H_\infty= 1] =
1$, defining $K := H$ and $L := 1$ would suffice for the purposes of
Theorem \ref{thm:doleanssharpened}. However, it might happen that
$\mathbb{P}[H_\infty> 1] > 0$ as is illustrated in Example \ref
{exa:coolstuff}. In this case, we shall construct the pair $(K, L)$
from $H$.
Before going to the technical details, we shall provide some intuition
on the definition of $(K, L)$. For $t \in\mathbb R_+$, $Z_t + \Delta
H_t =
\mathbb{E}[H_\infty- H_{t-} \mid\mathcal{F}_t]$ is the expected
total remaining
``life'' of $H$ on $[t, \infty[$, conditional on $\mathcal{F}_t$; then,
formally, $dH_t / (Z_t + \Delta H_t)$ is the ``fraction of remaining
life spent'' at $t$. The equivalent ``fraction of remaining life
spent'' for $K$, assuming that $K_\infty= 1$, would be $dK_t / (1 -
K_{t-})$. We shall ask that $K$ formally satisfies $dK_t / (1 -
K_{t-}) = dH_t / (Z_t + \Delta H_t)$ for $t \in\mathbb R_+$. To get a
feeling of how $L$ should be defined, observe that $\Delta K = (1 - K_-
) \Delta H / (Z + \Delta H)$ implies that $(1 - K) / Z = (1 - K_-)
/ (Z + \Delta H)$; therefore, formally, $dK_t / (1 - K_t) = dH_t
/ Z_t$ holds for $t \in\mathbb R_+$. Since $H = \int_{[0, \cdot]} L_t
\,dK_t$ has to hold in view of property (3) in Theorem
\ref{thm:doleanssharpened}, we obtain $L (1 - K) = Z$, which will be the defining
equation for $L$ as long as $K < 1$. We shall use the previous
intuition to define the pair $(K, L)$ rigorously below.

We proceed with our development, first assuming that $\mathbb{P}[H_t <
H_\infty\mid\mathcal{F}_t] = 1$ holds for all $t \in\mathbb R_+$---later, this
assumption will be removed. Under the previous assumption on $H$, it is
straightforward to see that $Z > 0$ (and, since $Z$ is a
supermartingale, also $Z_- > 0$) holds. We define $K$ as the unique
solution of the stochastic integral equation
\[
K = H_0 + \int_{(0, \cdot]} \biggl(\frac{1 - K_{t-}}{Z_t + \Delta H_t} \biggr)
\,dH_t,
\]
the latter being the rigorous equivalent of ``$dK_t / (1 - K_t) =
dH_t / (Z_t + \Delta H_t)$.''
The solution to the last equation is given by
%
\begin{eqnarray} \label{eq:defnofK}
K &=& 1 - (1 - H_0) \exp\biggl(- \int_{(0, \cdot]} \frac{dH_t}{Z_{t} +
\Delta H_t} \biggr) \nonumber\\[-8pt]\\[-8pt]
&&{}\times\prod_{t \in(0, \cdot]} \biggl(\biggl(1 - \frac {\Delta H_t}{Z_t
+ \Delta H_t} \biggr) \exp\biggl(\frac{\Delta H_t}{Z_t + \Delta H_t} \biggr) \biggr),\nonumber
\end{eqnarray}
which is an adapted, nondecreasing process with $0 \leq K < 1$, the
latter strict inequality holding due to our assumption on $H$. Set $L
:= Z / (1 - K)$, which is well defined in view of $K < 1$; $L$ is
nonnegative and $L_0 = Z_0 / (1 - K_0) = (1 - H_0) / (1 - H_0) = 1$.
Actually, $L$ is a local martingale. To see this, first observe that a
use of (\ref{eq:defnofK}) in reciprocal form gives
\[
\frac{1}{1 - K} = \frac{1}{1 - H_0} + \int_{(0, \cdot]} {\frac{d
H_t}{(1 - K_{t-} ) Z_t}}.
\]
Then, the integration-by-parts formula gives
\begin{eqnarray*}
L &=& \frac{Z}{1 - K} = 1 + \int_{(0, \cdot]} \frac{dZ_t}{1 -
K_{t-}} + \int_{(0, \cdot]} Z_t \,d\biggl(\frac{1}{1 - K_{t}} \biggr) \\
&=& 1 + \int_{(0, \cdot]} \frac{dZ_t}{1 - K_{t-}} + \int_{(0,
\cdot]}
Z_t \frac{dH_t}{( 1 - K_{t-}) Z_t} \\
&=& 1 + \int_{(0, \cdot]} \frac{dM_t}{1 - K_{t-}} \\
&=& 1 + \int_{(0, \cdot]} L_{t-} \frac{dM_t}{Z_{t-}}.
\end{eqnarray*}
The above string of equalities gives that $L$ is a local martingale,
and that it is actually equal to the stochastic exponential of the
local martingale $\int_{(0, \cdot]} (dM_t / Z_{t-} )$.

Now, drop the simplifying assumption $\mathbb{P}[H_t < H_\infty\mid
\mathcal{F}_t]
= 1$ for all $t \in\mathbb R_+$. Then, $Z > 0$ is no longer necessarily
true and more care has to be given in the definition of $K$ and $L$.
For each $n \in\mathbb N$, consider the stopping time $\tau^n :=
\inf
\{t \in\mathbb R_+ \mid Z_t \leq1 / n \}$, and define the predictable
set $\Theta:= \bigcup_{n \in\mathbb N} \llbracket0 ,\tau^n
\rrbracket$. Then,
$\Theta\subseteq\{Z_- > 0 \}$. Furthermore, with $\tau^\infty:=
\inf\{t \in\mathbb R_+ \mid Z_{t-} = 0 \mbox{ or } Z_{t} = 0 \}$, we
have $\uparrow\lim_{n \to\infty}\tau^n = \tau^\infty$, as well
as $\llbracket\tau ^\infty, \infty \llbracket\, = \{Z = 0 \}
\supseteq\{H = H_\infty \}$.

Define $K$ via (\ref{eq:defnofK}), and observe that $K$ is well
defined: our division conventions imply that $Z / (Z + \Delta H) = 1$
on $\{Z = 0 \}$, in view of the fact that $H$ is constant on $\{Z = 0 \}
$. It is clear that $K$ is adapted, right-continuous, nondecreasing
and $0 \leq K \leq1$. Furthermore, $K = \int_{[0, \cdot]} \mathbb{I}
_{\Theta_t} \,dK_t$ and $\Theta\subseteq\{K_- < 1 \}$. We shall
also consider the nonnegative local martingale $L$ that formally
satisfies $dL_t / L_{t-} = dM_t / Z_{t-}$ for $t \in\mathbb R_+$;
some care has to be given in defining $L$, since $Z_-$ might become
zero. Observe that $1 + \Delta M / Z_{-} = (Z + \Delta H) / Z_- \geq0$
holds on $\llbracket0, \tau^n \rrbracket$ for all $n \in\mathbb N$.
As $Z_- \geq1/n$
on $\llbracket0, \tau^n \rrbracket$, we can define a process $L^n$
as the stochastic
exponential of $\int_{(0, \tau^n \wedge\cdot]} (dM_{t} / Z_{t-})$.
Then, $L^n$ is a nonnegative local martingale, and $L^{n+1} = L^{n}$
holds on $\llbracket0, \tau_{n} \rrbracket$ for all $n \in\mathbb
N$. As
$(L^n_{\tau
^n})$ is a discrete-time nonnegative local martingale, $L_{\tau^\infty}
:= \lim_{n \to\infty}L^n_{\tau^n}$ $\mathbb{P}$-a.s. exists in
$\mathbb R_+$. It follows
that we can define a process $L$ such that $L = L^n$ on $\llbracket0,
\tau ^n \rrbracket$ for each $n \in\mathbb N$ and $L = L_{\tau
^\infty}$ on $\llbracket\tau^\infty, \infty \llbracket$. Note
that $L = \int_{(0, \cdot]} \mathbb{I}
_{\Theta
_t} \,dL_t = 1 + \int_{[0, \cdot]} \mathbb{I}_{\Theta_t} (L_{t-} / Z_{t-})
\,dM_t$. By the Ansel--Stricker theorem (see \cite{MR1277002}), $L$,
being a nonnegative process that is the stochastic integral of the
martingale $M$, is a local martingale. As $\Theta\subseteq\{K_- < 1 \}
$, $L = \int_{(0, \cdot]} \mathbb{I}_{\Theta_t} \,dL_t$ implies that $L
= \int_{[0, \cdot]} \mathbb{I}_{\{K_t < 1 \}} \,dL_t$. Furthermore,\vspace*{1pt} since
$\llbracket0, \tau^n \llbracket\ \subseteq\{L > 0 \}$ and $\{L_{\tau
^n} = 0 \} =
\{\Delta M_{\tau^n} / Z_{\tau^n -} = -1 \} = \{Z_{\tau^n} + \Delta
H_{\tau^n} = 0 \} = \{Z_{\tau^n} = 0, \Delta H_{\tau^n} = 0 \} =
\{\tau^n = \tau^\infty, \Delta H_{\tau^\infty} = 0 \}$ holds for
all $n
\in\mathbb N$, $K = \int_{[0, \cdot]} \mathbb{I}_{\Theta_t} \,dK_t$
implies $K = \int_{[0, \cdot]} \mathbb{I}_{\{L_t > 0 \}}
\,dK_t$.\vspace*{1pt}

With the above definitions, we shall establish that $L (1 - K) = Z$.
This result has already been obtained in a special case; we shall
utilize an approximation argument to show that it holds in general. For
any $\epsilon\in\mathbb R_+$, define the adapted, nonnegative,
nondecreasing and right-continuous process $H^\epsilon$ via
$H^\epsilon
_t = (H_t + \epsilon(1 - \exp(-t)) ) / (1 + \epsilon)$ for $t \in
\mathbb R
_+$. Then, for all $\epsilon\in\mathbb R_{++}$, $\mathbb
{E}[H^\epsilon
_\infty
] =1$, as well as $\mathbb{P}[H^\epsilon_t < H^\epsilon_\infty\mid
\mathcal{F}
_t] =
1$ holds for all $t \in\mathbb R_+$. Let $M^\epsilon$, $Z^\epsilon$,
$K^\epsilon$ and $L^\epsilon$ be the equivalents of the processes $M$,
$Z$, $K$ and $L$ defined with $H^\epsilon$ in place of $H = H^0$. Then,
$L^\epsilon(1 - K^\epsilon)= Z^\epsilon$ holds for all $\epsilon\in
\mathbb R_{++}$. It is straightforward to check that $Z^\epsilon_t =
(Z_t +
\epsilon\exp(-t)) / (1 + \epsilon)$, for all $t \in\mathbb R_+$; in
particular, $|Z^\epsilon- Z| \leq\epsilon(1 + Z) / (1 + \epsilon)$.
In view of $\mathbb{P}[\sup_{t \in\mathbb R_+} Z_t < \infty] = 1$,
we obtain
$\mathbb{P}[{\lim_{\epsilon\downarrow0} \sup_{t \in\mathbb R_+}}
|Z^\epsilon_t
- Z_t| = 0 ] = 1$. We shall also show the corresponding convergence of
$K^\epsilon$ to $K$ and $L^\epsilon$ to $L$ on every stochastic
interval $\llbracket0, \tau^n \rrbracket$, $n \in\mathbb N$. Define
a function
$\lambda\dvtx\mathbb R\mapsto\mathbb R_+ \cup\{\infty \}$ via $\lambda
(x) =
x - \log(1 + x)$ for $x \in\ ]{-}1, \infty[$ and $\lambda(x) = \infty$ for
$x \in\ ] {-} \infty, -1]$. Note that $0 \leq\lambda(a x) \leq
\lambda
(x)$ holds for all $x \in\mathbb R$ and $a \in[0,1]$, which will be used
in the limit theorems that will follow. Further, let $\mu^H$ be the
\textit{jump measure} of $H$, that is, the random counting measure on
$\mathbb R_+ \times\mathbb R$ defined via $\mu^H ((0,\cdot] \times
E) :=\sum
_{t \in(0, \cdot]} \mathbb{I}_{E \setminus\{0\}}(\Delta H_t)$ for $E
\subseteq\mathbb R$. A use of (\ref{eq:defnofK}), coupled with
straightforward algebra, allows us to write
\begin{eqnarray*}
1 - K^\epsilon_{\cdot\wedge\tau^n} &=& \frac{1 - H_0}{1 + \epsilon}
\exp\biggl(- \int_{(0, \cdot\wedge\tau^n]} \frac{dH_t}{Z_{t} + \Delta
H_t + \epsilon\exp(-t)} \\
&&\hspace*{55.95pt}{} - \epsilon\int_{(0, \cdot\wedge \tau
^n]} \frac{ \exp(-t) \,dt}{Z_{t} + \Delta H_t + \epsilon\exp(-t)} \biggr)
\\
&&{}\times
\exp\biggl(- \int_{(0, \cdot\wedge\tau^n] \times\mathbb R} \lambda
\biggl(\frac{x}{Z_{t} + \Delta H_t + \epsilon\exp(-t)} \biggr) \mu^H [dt, dx] \biggr).
\end{eqnarray*}
By straightforward applications of the monotone convergence theorem as
$\epsilon\downarrow0$ on the above equality, we obtain $\mathbb
{P}[\lim
_{\epsilon\downarrow0} \sup_{t \in[0, \tau^n]} |K^\epsilon_t - K_t|
= 0 ] = 1$ for all \mbox{$n \in\mathbb N$}. Furthermore, note that
$M^\epsilon
= (M + \epsilon) / (1 + \epsilon) $; therefore, $L^\epsilon$ is the
stochastic exponential of
\[
\int_{(0, \cdot]} \frac{dM^\epsilon_t}{Z^\epsilon_{t-}} = \int_{(0,
\cdot]} \frac{dM_t}{Z_{t-} + \epsilon\exp(-t)}.
\]
Let $\cMM:= [M, M] - \sum_{t \in[0, \cdot]} |\Delta M_t|^2$ be the
continuous part of the quadratic variation of $M$, and $\mu^M$ being
the jump measure of $M$ defined as $\mu^H$ before with ``$H$'' replaced
by ``$M$'' throughout. Using the definition of the stochastic
exponential, we obtain
\begin{eqnarray*}
L^\epsilon_{\cdot\wedge\tau^n} &=& \exp\biggl(\int_{(0, \cdot \wedge
\tau^n]} \frac{dM_t}{Z_{t-} + \epsilon\exp(-t)} - \frac{1}{2}
\int _{(0, \cdot\wedge\tau^n]} \frac{d\cMM_t}{\vert Z_{t-} +
\epsilon \exp(-t) \vert^2 } \biggr) \\
&&{}\times\exp\biggl(- \int_{\mathbb R\times(0, \cdot\wedge\tau^n]}
\lambda\biggl(\frac{x}{Z_{t-} + \epsilon\exp(-t)} \biggr) \mu^M [dx, dt] \biggr).
\end{eqnarray*}
The dominated theorem for stochastic integrals and the monotone
convergence theorem for ordinary Lebesgue integrals\vspace*{2pt} give $\mathbb
{P}[{\lim
_{\epsilon\downarrow0} \sup_{t \in[0, \tau^n]}} |L^\epsilon_t - L_t|
= 0] = 1$ for all $n \in\mathbb N$. It follows that $L (1 - K) = Z$
holds on $\Theta= \bigcup_{n \in\mathbb N} \llbracket0 ,\tau^n
\rrbracket$. As
$L =
\int_{[0, \cdot]} \mathbb{I}_{\Theta_t} \,dL_t$, $K = \int_{[0,
\cdot]}
\mathbb{I}_{\Theta_t} \,dK_t$ and\vspace*{1pt} $Z = \int_{[0, \cdot]} \mathbb{I}
_{\Theta_t}\,
dZ_t$, we obtain that $L (1 - K) = Z$ identically holds.

We have thus established that properties (1), (2), (3) and (4) of
Theorem \ref{thm:doleanssharpened} are satisfied by the pair $(L, K)$
that was constructed. Since $L (1 - K) = Z$ and $Z_\infty= 0$, the
set-inclusion $\{L_\infty> 0 \} \subseteq\{K_\infty= 1 \}$ is apparent.
\begin{rem} \label{rem:Lasmultdecomp}
When $H$ has continuous paths, $K$ has continuous paths as well---in
particular, $K$ is predictable. The formula $Z = L(1 - K)$ then implies
that $L$ coincides with the local martingale that appears in the
multiplicative decomposition of the nonnegative supermartingale $Z$.
\end{rem}
\begin{exa} \label{exa:coolstuff}
On $(\Omega, (\mathcal{F}_t)_{t \in\mathbb R_+}, \mathbb{P})$, let
$L$ be any nonnegative local martingale with $L_0 =
1$, $\Delta L \leq0$ and $L_\infty= 0$. Define $L^* = \max_{t \in[0,
\cdot]} L_t$; since $\Delta L \leq0$, $L^*$ is continuous. Define also
the nonnegative, nondecreasing, continuous and adapted process $H :=
\log(L^*)$, as well\vspace*{1pt} as $p$ via $\int_{\Omega\times\mathbb R_+} V \,dp
= \mathbb{E}[\int_{\mathbb R_+} V_t \,dH_t ]$ for all nonnegative
optional process $V$.
It is well known that $H_\infty= \log(L^*_\infty)$ has the standard
exponential distribution [see also (\ref{eq:doob'smaximal}) later
on]; therefore, $\mathbb{P}[H_\infty> 1] > 0$, $\mathbb{E}[H_\infty
] = 1$, and
$p$ is a unit-measure optional measure. Define $K := 1 - 1 / L^*$,
which is continuous, adapted, nondecreasing and satisfies $0 \leq K <
1$. Then,
\[
\int_{[0, \cdot]} L_t \,dK_t = \int_{[0, \cdot]} \frac{L_t}{|L^*_t|^2}
\,dL^*_t = \int_{[0, \cdot]} \frac{1}{L^*_t} \,dL^*_t = \log
(L^*) = H,
\]
where the second equality follows from the fact that the random measure
on $\mathbb R_+$ that is generated by the nondecreasing continuous process
$L^*$ is carried by the random set $\{t \in\mathbb R_+ \mid L_t =
L^*_t \}$. It follows that $(L, K)$ is actually a canonical
representation pair for $p$. Of course, it may happen that $L$ is a
strict local martingale; for example, $L$ could be the reciprocal of a
three-dimensional Bessel process starting from one. Observe also that
$\mathbb{P}
[K_\infty< 1] = \mathbb{P}[L^*_\infty< \infty] = 1$.
\end{exa}

\subsection{\texorpdfstring{Num\'{e}raire-invariant preferences on consumption
streams}{Numeraire-invariant preferences on consumption streams}}

Define $\Pi$ to be the class of measures on $(\Omega\times\mathbb
R_+, \mathcal{O})$ with unit
mass that are equivalent to some representative $\overline{p} \in\Pi$.
Then, let $\mathcal{I}$ be the class of all adapted, right-continuous,
nonnegative and nondecreasing processes $F$ satisfying the following
property: if $A \in\mathcal{O}$ is $\Pi$-null, $\int_{[0, \cdot]}
\mathbb{I}
_{A_t} \,dF_t$ is an evanescent process. The processes in $\mathcal{I}$ model
all cumulative consumption streams that an agent could potentially
choose from; if $A \in\mathcal{O}$ is $\Pi$-null, the agent gives no
consumption value on $A$, and therefore will not consume there. The
following result gives a convenient characterization of the set
$\mathcal{I}$.
\begin{prop} \label{prop:charactofincreset}
Fix $p \in\Pi$ with canonical representation pair $(L, K)$. Then,
$\mathcal{I}
$ is the class of all finite processes $\int_{[0, \cdot]} a_t \,dK_t$,
where $a$ ranges though the nonnegative optional processes.
\end{prop}
\begin{pf}
Let $p \in\Pi$, and let $(L, K)$ be a pair of adapted c\`{a}dl\`{a}g
processes satisfying properties (1), (2), (3) and (4) of Theorem
\ref{thm:doleanssharpened}. Let also $H := \int_{[0, \cdot]} L_t
\,d
K_t$, so that $\int_{\Omega\times\mathbb R_+} V \,dp = \mathbb
{E}[\int_{\mathbb R_+} V_t \,dH_t ]$ holds for all nonnegative optional
process $V$. For $A \in
\mathcal{O}$, $p[A] = 0$ if and only if $\int_{[0, \cdot]} \mathbb{I}_{A_t}
d
H_t$ is evanescent.

By Theorem V.5.14 of \cite{MR1219534}, for all $F \in\mathcal{I}$ there
exists an nonnegative optional process $b$ such that $F = \int_{[0,
\cdot]} b_t \,dH_t$. Letting $a := b L$, we have $F = \int_{[0,
\cdot
]} a_t \,dK_t$.

Now, let $A \in\mathcal{O}$. We have $p[A] = 0$ if and only if $\int_{[0,
\cdot]} \mathbb{I}_{A_t} L_t \,dK_t$ is evanescent. As $L$ is a
nonnegative local martingale, this is equivalent to saying that $\int
_{[0, \cdot]} \mathbb{I}_{A_t} \mathbb{I}_{\{L_t > 0 \}} \,dK_t$ is
evanescent. Since\vspace*{1pt} $K = \int_{[0 ,\cdot]} \mathbb{I}_{\{L_{t} > 0 \}}
\,d
K_t$, this is further equivalent to saying that $\int_{[0, \cdot]}
\mathbb{I}_{A_t} \,dK_t$ is evanescent. To\vspace*{1pt} recapitulate, $A \in\mathcal
{O}$ is
$\Pi$-null if and only if $\int_{[0, \cdot]} \mathbb{I}_{A_t} \,dK_t$ is
evanescent. We then have $K \in\mathcal{I}$, and therefore, $\int_{[0,
\cdot
]} a_t \,dK_t$ also belongs to $\mathcal{I}$ for each nonnegative optional
process $a$ such that the last integral is nonexploding in finite
time. This completes the argument.
\end{pf}
\begin{rem} \label{rem:uniquenessnotused}
The essential uniqueness of a canonical representation pair $(L, K)$
for $p \in\Pi$, which has not been established yet, was \textit{not}
used in the proof above. Just the existence of a pair $(L, K)$ that
satisfies properties (1), (2), (3) and (4) of Theorem \ref
{thm:doleanssharpened} was utilized, which was shown in Section
\ref{subsec:proofofthm:doleanssharpened}.
\end{rem}

In view of the previous result, for $p \in\Pi$, and with $(L, K)$ a
canonical representation pair for $p$, each $F \in\mathcal{I}$ can be written
as $F = \int_{[0, \cdot]} \partial^{F|K}_t \,dK_t$. Then, for $F
\in
\mathcal{I}$ and $G \in\mathcal{I}$ we define
%
\begin{equation} \label{eq:dfnofR-Nderiv}
\frac{dF}{dG} := \frac{\partial^{F|K}}{\partial^{G|K}},
\end{equation}
where once again we are using the conventions on division discussed in
the first paragraph of Section \ref{subsec:probnotanddefn}.
If $p' \in\Pi$ has canonical representation pair $(L', K')$, then,
since $K \in\mathcal{I}$ and $K' \in\mathcal{I}$, we have $\partial
^{K'|K} > 0$ and
$\partial^{K|K'} > 0$ holding $\Pi$-a.e., as well as $\partial
^{F|K'} =
\partial^{F|K} \partial^{K|K'}$, $\Pi$-a.e., for all $F \in\mathcal{I}$.
Therefore, the definition of $\partial F / \partial G$ in
(\ref{eq:dfnofR-Nderiv}) does not depend on the choice of $p \in\Pi$.

For $p \in\Pi$ with canonical representation pair $(L, K)$, and all $F
\in\mathcal{I}$ and $G \in\mathcal{I}$, we define
%
\begin{equation} \label{eq:relratecons}
\mathsf{rel}_p ( F | G ) := \int_{\Omega\times\mathbb R_+} \biggl(\frac
{dF}{dG} \biggr) \,dp
- 1 =
\mathbb{E}\biggl[\int_{\mathbb R_+} \biggl(\frac{\partial^{F|K}_t}{\partial
^{G|K}_t} \biggr) L_t \,dK_t \biggr] - 1
\end{equation}
and the corresponding preference relation $\preccurlyeq_p$ on
$\mathcal{I}$ via $F
\preccurlyeq_p G \Longleftrightarrow\mathsf{rel}_p ( F | G ) \leq0$
for all $F \in
\mathcal{I}$ and $G \in\mathcal{I}$.

Such preference relations can be seen to stem from axiomatic
foundations, just as in the static case that is presented in Theorem
\ref{thm:converse}. Since the details of such generalization are
straightforward, we shall not delve into them here. Rather, we shall
focus on novel features appearing in a dynamic environment.
\begin{rem} \label{rem:subjectiveviews2}
Recall the discussion in Remark \ref{rem:subjectiveviews1}. Let $p
\in\Pi$ with canonical representation pair $(L, K)$, and suppose that
$L$ is the density process of a probability $\mathbb{Q}$ with respect to
$\mathbb{P}$. Then, $\mathbb{Q}[K_\infty= 1] = 1$, and
\[
\mathsf{rel}_p ( F | G ) = \mathbb{E}_\mathbb{Q}\biggl[\int_{\mathbb R_+}
\biggl(\frac{dF_t - d G_t }{dG_t} \biggr) \,dK_t \biggr] = \mathbb{E}_\mathbb{Q}\biggl[\int
_{\mathbb R_+} \biggl(\frac {\partial^{F|K}_t - \partial^{G|K}_t
}{\partial^{G|K}_t} \biggr) \,dK_t \biggr]
\]
holds for all $F \in\mathcal{I}$ and $G \in\mathcal{I}$. We
interpret $\mathbb{Q}$ as
the \textit{subjective views} of an agent and $K$ as the agent's
\textit{consumption clock}. As was described in Example \ref{exa:coolstuff},
$L$ might fail to be the density process of a probability $\mathbb{Q}$ with
respect to $\mathbb{P}$, and $\mathbb{P}[K_\infty= 1] = 1$ might
fail. We still
``loosely'' interpret $L$ as subjective views and $K$ as consumption clock.
\end{rem}

\subsection{The investment--consumption problem} \label{subsec:inv-cons}

The canonical representation pair for an optional measure with unit
mass allows for a very satisfactory solution to an agent's
investment--consumption problem.

\subsubsection{Pure investment} \label{subsubsec:pureinvestment}
Henceforth, $S = (S^i)_{i=1, \ldots, d}$ will be a vector-valued
semimartingale. For each $i \in\{1, \ldots, d \}$, $S^i$ should be
thought as representing the discounted, with respect to some baseline
security, price of a liquid asset traded in the market, satisfying $S^i
> 0$ and $S^i_- > 0$.

Consider a set-valued process $\mathfrak{K}\dvtx \Omega\times\mathbb R_+
\mapsto
2^{\mathbb R^d} \setminus\{\varnothing \}$, where $2^{\mathbb R^d}$
denotes the
powerset of $\mathbb R^d$, which will represent constraints imposed on the
agent on the percentage of capital-at-hand invested in the liquid
assets. The last set-valued process is assumed to satisfy some natural
properties; namely, $\mathfrak{K}(\omega, t)$ is convex and closed
for all
$(\omega, t) \in\Omega\times\mathbb R_+$, $\mathfrak{K}$ is
predictable, in the sense that the
set $\{(\omega, t) \in\Omega\times\mathbb R_+\mid\mathfrak
{K}(\omega, t) \cap A \neq \varnothing \}$ is predictable for all
closed $A \subseteq\mathbb R^d$, and
finally $\mathfrak{K}$ large enough as to contain all investments that produce
zero wealth. Under a simple nonredundancy condition on the liquid
assets, the last requirement simply reads $0 \in\mathfrak{K}(\omega,
t)$ for
all $(\omega, t) \in\Omega\times\mathbb R_+$. More precise
information about these
requirements can be found in \cite{MR2335830}.

Starting with capital $x \in\mathbb R_+$, and investing according to some
$d$-dimensional, predictable strategy $\theta$ representing the number
of liquid assets held in the portfolio, an economic agent's discounted
wealth is given by
%
\begin{equation} \label{eq:wealthprocess}
X^{x, \theta} = x + \int_{[0, \cdot]} \langle\theta_t , dS_t
\rangle.
\end{equation}
We define
\begin{eqnarray*}
&&\mathcal{X}(x) := \bigl\{X^{x, \theta} \mid X^{x, \theta} \mbox{ is
defined in } (\ref{eq:wealthprocess}), X^{x, \theta} \geq0,\\
&&\hspace*{53.4pt}\hspace*{17.8pt} \mbox{and } \{ (\theta^i S^i_-)_{i=1, \ldots, d} \in X^{x, \theta}_-
\mathfrak{K}\} \mbox{ is } \Pi\mbox{-full} \bigr\}.
\end{eqnarray*}
The elements of $\mathcal{X}(x)$ are \textit{pure-investment} outcomes, starting
with initial capital $x \in\mathbb R_+$. We also set $\mathcal{X}=
\bigcup_{x
\in
\mathbb R_+} \mathcal{X}(x)$. The next result regards the viability of
the market.
Its validity follows from Theorem 4.12 in \cite{MR2335830} coupled with
a localization argument; its straightforward proof is omitted.
\begin{theorem} \label{thm:numeraireunderlocalprobchanges}
With the above notation, the following two conditions are equivalent:
\begin{enumerate}[(2)]
\item[(1)] For all $t \in\mathbb R_+$, the set $\{X_t \mid X \in\mathcal
{X}(1) \}
\subseteq{\mathbb{L}^0_+}$ is bounded.
\item[(2)] For any nonnegative local martingale $L$ with $L_0 = 1$, there
exists $\widehat{X}^L \in\mathcal{X}(1)$ such that:
\begin{enumerate}[(a)]
\item[(a)] $L(X / \widehat{X}^L)$ is a supermartingale for all $X \in
\mathcal{X}$.\vspace*{1pt}
\item[(b)] $\int_{(0 ,\cdot]} \mathbb{I}_{\{L_{t-} = 0 \}}\, d\widehat
{X}^L_t$ is an
evanescent process.
\end{enumerate}
With the above specifications, $\widehat{X}^L$ is unique up to
indistinguishability.
\end{enumerate}
Under any of the above equivalent conditions, we have $\mathcal{X}(0)
= \{0 \}$.
\end{theorem}
\begin{rem}
In the spirit and notation of the discussion of Remark \ref
{rem:subjectiveviews2}, and if $L$ is the density process of a probability
$\mathbb{Q}$ with respect to $\mathbb{P}$, the process $\widehat
{X}^L$ of Theorem
\ref{thm:numeraireunderlocalprobchanges} above is simply the
\textit{num\'{e}raire portfolio} under $\mathbb{Q}$ (see
\cite{LONG,MR1849424,MR2335830}). According to Theorem \ref
{thm:numeraireunderlocalprobchanges}, the equivalent of the num\'{e}raire
portfolio when the
``views'' of the agent are given by $L$ exists even in cases where $L$
is a strict local martingale and does not stem from a change of probability.
\end{rem}

\subsubsection{Investment and consumption}

We now introduce agent's consumption. For $x \in\mathbb R_+$, a
\textit{consumption stream} $C \in\mathcal{I}$ will be called \textit{financeable}
starting from capital $x \in\mathbb R_+$ if there exists a predictable,
$d$-dimensional and $S$-integrable $\eta$ with the property that $X^{x,
\eta, C} := X^{x, \eta} - C$ is such that $X^{x, \eta, C} \geq0$ and
$ \{ (\eta^i S^i_-)_{i=1, \ldots, d} \in X^{x, \eta, C}_- \mathfrak{K}
\}
$ is $\Pi$-full. The class of all consumption streams that can be
financed starting from $x \in\mathbb R_+$ will be denoted by $\mathcal
{C}(x)$. It
is straightforward that $\mathcal{C}(x) = x \mathcal{C}(1)$ for $x
\in\mathbb R_{++}$.
Furthermore, under any of the equivalent conditions of Theorem
\ref{thm:numeraireunderlocalprobchanges}, $\mathcal{C}(0) = \{0 \}$ holds.

For the solution to the agent's optimal investment--consumption problem
that will be presented in Theorem \ref{thm:invest-cons} below, a
``multiplicative'' representation for elements of $\mathcal{C}(x)$, $x
\in
\mathbb R
_+$ will turn out to be more appropriate. To begin with, let ${\mathcal{I}(1)}$
be the set of all $F \in\mathcal{I}$ with $F_\infty\leq1$; observe that
${\mathcal{I}(1)}$ corresponds to the set $\mathcal{C}(1)$ if $S =
0$, that is, if there
are no investment opportunities. For $x \in\mathbb R_{++}$, let $C \in
\mathcal{C}
(x)$, and let $\eta$ be a strategy that finances $C$. Then, we can
write $X^{x, \eta, C} = X^{x, \theta} (1 - F)$, where $F \in
{\mathcal{I}(1)}$
formally satisfies $dF_t / (1 - F_t) = dC_t / X^{x, \theta, C}_t$
[in other words, $dF_t / (1 - F_t)$ is the rate of consumption
relative to the capital-at-hand], and $\theta:= (1 / (1 - F_-)) \eta
$. Note also that $ \{ (\theta^i S^i_-)_{i=1, \ldots, d} \in X^{x,
\theta}_- \mathfrak{K}\} = \{ (\eta^i S^i_-)_{i=1, \ldots, d} \in
X^{x, \eta, C}_- \mathfrak{K}\}$, which is $\Pi$-full, and therefore
$X^{x, \theta} \in\mathcal{X}(x)$. Conversely, start with $X^{x,
\theta} \in
\mathcal{X}
(x)$ and $F \in{\mathcal{I}(1)}$ and define $C := \int_{[0, \cdot]} X^{x,
\theta}_{t} \,dF_t$ and $\eta:= (1 - F_-) \theta$. Then, $X^{x,
\eta
, C} = X^{x, \theta} (1 - F)$ and $ \{ (\eta^i S^i_-)_{i=1,
\ldots,
d} \in X^{x, \eta, C}_- \mathfrak{K}\} = \{ (\theta^i S^i_-)_{i=1,
\ldots, d} \in X^{x, \theta}_- \mathfrak{K}\}$, which is $\Pi
$-full. Under
any of the equivalent conditions of Theorem \ref
{thm:numeraireunderlocalprobchanges}, since $\mathcal{X}(0) = \{0 \} =
\mathcal{C}
(0)$, an alternative
equivalent description the class of financeable consumption streams
starting from capital $x \in\mathbb R_+$ is
%
\begin{equation} \label{eq:consumptionset}
\mathcal{C}(x) = \biggl\{\int_0^\cdot X_t \,dF_t \Bigm| X \in\mathcal{X}(x)
\mbox { and } F \in{\mathcal{I}(1)} \biggr\}.
\end{equation}
\begin{theorem} \label{thm:invest-cons}
Let $p \in\Pi$ with canonical representation pair $(L, K)$. Assume any
of the equivalent conditions of Theorem \ref
{thm:numeraireunderlocalprobchanges}, and let $\widehat{X}^L \in
\mathcal{X}(1)$ be
defined as in the latter
result. Fix $x \in\mathbb R_{+}$ and define $\mathcal{C}(x)$ via
(\ref{eq:consumptionset}). Then, with $\widehat{C} := x \int_{[0, \cdot]}
\widehat{X}
^L_t \,dK_t \in\mathcal{C}(x)$, $C \preccurlyeq_p \widehat{C}$ holds
for all $C
\in
\mathcal{C}(x)$.
\end{theorem}
\begin{pf}
For $x \in\mathbb R_{++}$, fix $X \in X(x)$ and $F \in{\mathcal
{I}(1)}$ and let $C
= \int_0^\cdot X_t \,dF_t = \int_0^\cdot X_t \partial^{F|K}_t \,dK_t$.
Let $N := (1 / x) L (X / \widehat{X}^L)$. Then, recalling that
\[
\widehat{C}:= x
\int_{[0, \cdot]} \widehat{X}^L_t \,dK_t,
\]
we have
\[
\mathsf{rel}_p ( C | \widehat{C} ) = \mathbb{E}\biggl[\int_{\mathbb R_+}
\frac{X_t \partial ^{F|K}_t}{x \widehat{X}^L_t} L_t \,dK_t \biggr] - 1 =
\mathbb{E}\biggl[\int_{\mathbb R_+} N_t \,dF_t \biggr] - 1.
\]
For any finite stopping time $\tau$, and in view of $N_0 = 1$, one has
\begin{eqnarray*}
\int_{[0, \tau]} N_t \,dF_t - 1 &=& N_\tau F_\tau- N_0 - \int_{[0,
\tau
]} F_{t-} \,dN_t \leq N_\tau- N_0 - \int_{[0, \tau]} F_{t-} \,dN_t \\
&=&
\int_{(0, \tau]} (1 - F_{t-}) \,dN_t.
\end{eqnarray*}
Pick an increasing sequence $(\tau^n)_{n \in\mathbb N}$ of stopping
times that $\mathbb{P}$-a.s. converges to infinity and is such that
$\mathbb{E}
[ \sup_{t \in[0, \tau^n]} N_t ] < \infty$ for all $n \in
\mathbb N$. Then, $\mathbb{E}[ \int_{(0 ,\tau^n]} (1 - F_{t-}) \,dN_t
] \leq0$ hold for all $n \in\mathbb N$ because $N$ is a
nonnegative supermartingale and $0 \leq F \leq1$. Therefore,
\[
\mathsf{rel}_p ( C | \widehat{C} ) = \lim_{n \to\infty}\mathbb
{E}\biggl[\int_{[0, \tau^n]} N_t \,d F_t \biggr] - 1
\leq\mathop{\lim\sup}_{n \to\infty}\mathbb{E}\biggl[\int_{(0, \tau
^n]} (1 - F_{t-}) \,dN_t \biggr]
\leq0,
\]
which completes the proof.
\end{pf}

The result of Theorem \ref{thm:invest-cons} describes how an agent
with num\'{e}raire-invariant preferences generated by $p$ will dynamically
invest and consume in an optimal manner. The canonical representation
pair $(L, K)$ of $p$ conveniently separates the investment and
consumption problems. The optimal strategy, when described in
proportions of wealth invested in the assets, is completely
characterized by $L$; indeed, these proportions will be the same as the
ones held in the portfolio that results in the pure-investment wealth
$\widehat{X}^L$. On the other hand, the optimal consumption in an infinitesimal
interval around $t \in\mathbb R_+$ relative to the capital-at-hand is
$d
K_t / (1 - K_t)$, which solely depends on $K$.

As can be seen from its proof, the validity of Theorem \ref
{thm:numeraireunderlocalprobchanges} goes well beyond the framework of
investing in a market with certain finite number of liquid assets. All
that is needed is a class of nonnegative ``wealth'' processes
$(\mathcal{X} (x) )_{x \in\mathbb R_+}$ with $\mathcal{X}(x) = x
\mathcal{X}(1)$ for $x \in\mathbb R_+$, such
that statement (2) of Theorem \ref{thm:numeraireunderlocalprobchanges}
holds; in other words, the crucial element is the existence of
a num\'{e}raire portfolio under the ``local change in probability''
with the
local martingale $L$ acting as a ``density process.'' The computational
advantage of assuming a semimartingale $S$ that generates the wealth
processes is that the process $\widehat{X}^L$ appearing in Theorems
\ref{thm:numeraireunderlocalprobchanges} and~\ref{thm:invest-cons}
can be completely described by the use of the triplet of predictable
characteristics (see \cite{MR1943877}) of the $(1+d)$-dimensional
process $(L, S)$. The formulas appear in \cite{MR1970286}, where the
closely-related problem of log-utility consumption maximization under a
random clock is treated. Nevertheless, in the latter paper, the authors
did not utilize the canonical representation pair in the solution; for
this reason, unless the consumption clock is deterministic, it is not
apparent that the two aspects of investment and consumption can be
separated, as was previously pointed out.
\begin{rem} \label{rem:purecons}
Theorem \ref{thm:invest-cons} solves in particular the \textit{pure
consumption problem}. Assume that an agent stats with a unit of
account, has no access in a market and needs to choose how this unit of
account will be consumed throughout time. This is modeled by setting
$\mathcal{C}
(1) = {\mathcal{I}(1)}$. Let $p \in\Pi$ with $(L, K)$ be its canonical
representation pair. Then, $F \preccurlyeq_p K$ holds for all $F \in
{\mathcal{I}(1)}
$. Note that the optimal solution does \textit{not} depend on $L$, in par
with the discussion that followed Theorem~\ref{thm:invest-cons}.

In fact, the same consumption stream $K$ solves the optimization
problem for more general preference structures. Let $U \dvtx \mathbb R_{++}
\mapsto\mathbb R$ be a concave and nondecreasing function, and extend the
definition of $U$ by setting $U(0) = \lim_{x \downarrow0} U(x)$.
Consider a preference structure on ${\mathcal{I}(1)}$ with numerical
representation given via the utility functional
\[
{\mathcal{I}(1)}\ni F \mapsto\mathbb{U}(F) = \int_{\Omega\times
\mathbb R_+} U(\partial^{F|K} )
\,dp = \mathbb{E}\biggl[\int_{\mathbb R_+} U (\partial^{F|K}_t ) L_t \,dK_t \biggr],
\]
where we shall soon see that the above integrals are well defined, in
the sense that the positive part of the integrand is integrable. Let
$(\tau^n)_{n \in\mathbb N}$ be a localizing sequence such that
$\mathbb{E}
[\sup_{t \in[0, \tau^n]} L_t ] < \infty$ for all $n \in
\mathbb N$. Since
\begin{eqnarray*}
\int_{\Omega\times\mathbb R_+} \partial^{F|K} \,dp &=& \mathbb
{E}\biggl[\int_{\mathbb R_+} L_t \,dF_t \biggr] = \lim_{n \to\infty}\mathbb
{E}\biggl[\int_{[0, \tau^n]} L_t \,d F_t \biggr] \\
&=& \lim_{n \to\infty}\mathbb{E}\biggl[L_{\tau^n} F_{\tau^n} - \int
_{[0, \tau^n]} F_{t-} \,dL_t \biggr] \leq1,
\end{eqnarray*}
Jensen's inequality gives $\mathbb{U}(F) \leq U (\int_{\Omega\times
\mathbb R_+} \partial ^{F|K} \,dp ) \leq U(1) = \mathbb{U}(K)$.
Therefore, $K$ is the optimal
consumption plan.
\end{rem}

\subsection[Essential uniqueness of a canonical representation pair in
Theorem 2.1]{Essential uniqueness of a canonical representation pair in
Theorem \protect\ref{thm:doleanssharpened}}
\label{subsec:uniquenessofcanreppair}

Let $p \in\Pi$, and let $(L, K)$ and $(L', K')$ be two pairs of
processes having the properties (1), (2), (3) and (4) in Theorem
\ref{thm:doleanssharpened}. The equality $\int_{[0, \cdot] } L_t \,dK_t
= \int_{[0, \cdot] } L'_t \,dK'_t$ holds due to the uniqueness of
Dol\'{e}ans's representation of $p$.

Since $K \in{\mathcal{I}(1)}$ and $K' \in{\mathcal{I}(1)}$, Theorem
\ref
{thm:invest-cons} implies that $K \preccurlyeq_p K'$ and \mbox{$K'
\preccurlyeq_p K$}. (In
view of Remark \ref{rem:uniquenessnotused}, the result of Theorem
\ref{thm:invest-cons} does not assume uniqueness of canonical
representation pairs; therefore, there is no cyclic argument.) It
follows that $\partial^{K' | K} = 1$ holds $\Pi$-a.e., or, in other
words, that $K = K'$ in the sense that $K$ and $K'$ are indistinguishable.

Since $K = K'$, the equality $\int_{[0, \cdot] } L_t \,dK_t = \int
_{[0, \cdot] } L'_t \,dK'_t$ translates to $K L - \int_{[0, \cdot] }
K_{t-} \,dL_t = K L' - \int_{[0, \cdot] } K_{t-} \,dL'_t$. Let
$(\tau
^n)_{n \in\mathbb N}$ be a nondecreasing sequence of stopping times
such that, $\mathbb{P}$-a.s., $\uparrow\lim_{n \to\infty}\tau^n =
\infty$, as well as
$\mathbb{E}[ \sup_{t \in[0, \tau^n]} L_t ] < \infty$ and
$\mathbb{E}
[ \sup_{t \in[0, \tau^n]} L'_t ] < \infty$ holds for all $n
\in\mathbb N$. Then, $\mathbb{E}[K_{\tau\wedge\tau^n} L_{\tau
\wedge
\tau
^n}] = \mathbb{E}[K_{\tau\wedge\tau^n} L'_{\tau\wedge\tau^n}]$ holds
for all $n \in\mathbb N$ and stopping times $\tau$. Since $L$, $L'$ and
$K$ are all adapted c\`{a}dl\`{a}g processes, it follows that $K L$ and $K
L'$ are indistinguishable. This, coupled with the fact that $L$ and
$L'$ are both local martingales, gives $\{K_\infty> 0 \} \subseteq
\{L_t = L'_t, \forall t \in\mathbb R_+ \}$.

\subsection{A random time-horizon investment problem}
\label{subsec:randomhorizon}

We retain all the notation from Section \ref{subsubsec:pureinvestment}
for the market description and the investment sets. We
shall also be assuming throughout that the market satisfies the
viability requirement of Theorem \ref
{thm:numeraireunderlocalprobchanges}. In particular, recall the
notation $\widehat{X}^L \in\mathcal{X}(1)$ from the
last result. We are interested in characterizing the equivalent of the
num\'{e}raire portfolio under $\mathbb{P}$, sampled at a random, \textit{not
necessarily stopping}, time. Here, by a \textit{random time} we simply
mean a $\mathbb R_{+}$-valued, $\mathcal{F}_\infty$-measurable random
variable $T$.
\begin{theorem} \label{thm:numunderrandomsampling}
For any random time $T$, define the measure $p = p^{T}$ on $(\Omega
\times\mathbb R_+,
\mathcal{O})$ via $\int_{\Omega\times\mathbb R_+} V \,dp = \mathbb
{E}[V_T]$ for all nonnegative
optional process $V$. Since $p [\Omega\times\mathbb R_+] = 1$ and
$p[A] = 0$ holds for
all evanescent $A \in\mathcal{O}$, let $(L, K)$ be the canonical
representation pair for $p$. Then, $\mathbb{E}[X_T / \widehat{X}^L_T
] \leq
X_0 / \widehat{X}^L_0 = X_0$ holds for all $X \in\mathcal{X}$.
\end{theorem}
\begin{pf}
For $X \in\mathcal{X}(1)$, define $C := \int_{[0, \cdot]} X_t \,dK_t$.
Define also
\[
\widehat{C} := \int_{[0, \cdot]} \widehat{X}^L_t
\,dK_t.
\]
Then,
$C \in\mathcal{C}(1)$, $\widehat{C} \in\mathcal{C}(1)$ and
$\mathbb{E}[\int _{\mathbb R_+} (\partial^{C | K }_t / \partial
^{\widehat{C} | K }_t ) L_t \,d K_t ] \leq1$. Therefore,
\begin{eqnarray*}
\mathbb{E}\biggl[\frac{X_T}{\widehat{X}^L_T} \biggr] &=& \int_{\Omega\times
\mathbb R_+} \biggl(\frac {X}{\widehat{X} ^L} \biggr) \,dp = \mathbb{E}\biggl[\int
_{\mathbb R_+} \biggl(\frac{X_t}{\widehat{X} ^L_t} \biggr) L_t \,dK_t \biggr] \\
&=& \mathbb
{E}\biggl[\int_{\mathbb R_+} \biggl(\frac{\partial^{C | K }_t}{\partial
^{\widehat{C} | K }_t} \biggr) L_t \,dK_t \biggr] \leq1.
\end{eqnarray*}
The result follows by simply noting that $\mathcal{X}(x) = x \mathcal
{X}(1)$ holds for
all $x \in\mathbb R_+$.
\end{pf}

The next result is a partial converse to Theorem \ref
{thm:numunderrandomsampling}, in the sense that the nonnegative local martingale
$L$ will be given and the random time $T$ will be constructed from $L$.
Recall that the \textit{jump process} of a process $L$ is defined via
$\Delta L_t = L_t - L_{t-}$ for all $t \in\mathbb R_+$.
\begin{theorem} \label{thm:locmarttoclock}
Let $L$ be a nonnegative local martingale with $L_0 = 1$, $\Delta L
\leq0$ and $L_\infty= 0$. Let $T$ be any random time with $L_T = \max
_{t \in\mathbb R_+} L_t$. Then, $\mathbb{E}[X_T / \widehat{X}^L_T ]
\leq X_0 /
\widehat{X}
^L_0 = X_0$ holds for all $X \in\mathcal{X}$.
\end{theorem}
\begin{pf}
The key to proving Theorem \ref{thm:locmarttoclock} is the
following version of Doob's maximal identity, which can be found for
example in Lemma 2.1 of \cite{MR2247846}: for all finite stopping times
$\tau$ and $\mathcal{F}_\tau$-measurable and nonnegative random variables
$\gamma
$, one has
%
\begin{equation} \label{eq:doob'smaximal}
\mathbb{P}\Bigl[\sup_{t \in[\tau, \infty)} L_t > \gamma\bigm|\mathcal
{F} _\tau \Bigr]
= \biggl(\frac{L_\tau}{\gamma} \biggr) \wedge1.
\end{equation}

The assumption $\Delta L \leq0$ implies that the nondecreasing process
$L^* :=\break \max_{t \in[0, \cdot]} L_t$ is continuous. Consider the
random times $T_{\sup} := \sup\{t \in\mathbb R_+ \mid L_t =
L^*_\infty \}$ and
$T_{\inf} := \inf\{t \in\mathbb R_+ \mid L_t = L^*_\infty \}$.
Obviously, $T_{\inf} \leq T \leq T_{\sup}$. A use of
(\ref{eq:doob'smaximal}) gives that for any finite stopping time $\tau$ we
have $\mathbb{P}
[ T_{\sup} > \tau\mid\mathcal{F}_\tau] = \mathbb{P}[ \sup_{t
\in
[\tau, \infty)} L_t \geq L^*_\tau\mid\mathcal{F}_\tau] = L_\tau/
L^*_\tau$, as well as the equality $\mathbb{P}[T_{\inf} > \tau \mid
\mathcal{F} _\tau ] = \mathbb{P}[ \sup_{t \in[\tau, \infty)} L_t
> L^*_\tau
\mid
\mathcal{F}_\tau] = L_\tau/ L^*_\tau$.

Define the measure $p^{T}$ on $(\Omega\times\mathbb R_+, \mathcal
{O})$ via $\int_{\Omega\times\mathbb R_+} V
\,dp^T = \mathbb{E}[V_T] =\break \mathbb{E}[ \int_{\mathbb R_+} V_t \,dH_t ]$
for nonnegative optional processes $V$, where $H$ is the dual optional
projection of the process $\mathbb{I}_{\llbracket T, \infty
\llbracket}$. Let $Z$ be the
nonnegative supermartingale such that $Z_t = \mathbb{E}[H_\infty- H_t
\mid\mathcal{F}_t] = \mathbb{P}[T > t \mid\mathcal{F}_t]$ holds
for all $t \in\mathbb R_+$.
Since $T_{\inf} \leq T \leq T_{\sup}$, it follows that $Z = L / L^*$.
In the notation of Theorem \ref{thm:doleanssharpened}, and according
to Remark \ref{rem:Lasmultdecomp}, $L$ is the local martingale in
the canonical representation pair of $p^T$. Then, it follows from
Theorem \ref{thm:numunderrandomsampling} that $\mathbb{E}[X_T /
\widehat{X}
^L_T] \leq X_0$ for all $X \in\mathcal{X}$.
\end{pf}

Let $S$ be a one-dimensional semimartingale that generates the
wealth-process class $\mathcal{X}$. Assume that $S > 0$, $\Delta S
\geq0$, $1 /
S$ is a local martingale and $\lim_{t \to\infty} S_t = \infty$. Define
$L = S_0 / S$, and let $T$ be any random time such that $S_T = \min_{t
\in\mathbb R_+} S_t$, that is, $L_T = \max_{t \in\mathbb R_+} L_t$.
It is
straightforward to see that $\widehat{X}^L = 1$ and $\widehat{X}^1 =
S / S_0 = 1 /
L$. In view of Theorem \ref{thm:locmarttoclock}, it follows that
$\mathbb{E}[X_T] \leq X_0$ for all $X \in\mathcal{X}$. In words, at
the random
time of the overall minimum of $S$, which is the time of the overall
minimum the num\'{e}raire portfolio, the whole market is at a
downturn. We
shall show below that the last fact is always true, regardless of
whether $S$ is a one-dimensional semimartingale with $1/S$ is a local
martingale or not. The next result adds yet one more remarkable fact to
the long list of optimality properties of the num\'{e}raire portfolio, with
the loose interpretation of the num\'{e}raire portfolio being an index of
market status.
\begin{theorem} \label{thm:moreimportanceofnumeraire}
Suppose that $\widehat{X}\equiv\widehat{X}^1 \in\mathcal{X}(1)$ is
such that $\Delta\widehat{X}
\geq
0$ and $\lim_{t \to\infty} \widehat{X}_t = \infty$. Let $T$ be any
random time
such that $\widehat{X}_T = \min_{t \in\mathbb R_+} \widehat{X}_t$.
Then, $\mathbb{E}[X_T]
\leq
X_0$ holds for all $X \in\mathcal{X}$.
\end{theorem}
\begin{pf}
Let $L := 1 / \widehat{X}$. Since $\widehat{X}\in\mathcal{X}(1)$,
$L_0 = 1$. Also,
$\Delta\widehat{X}\geq0$ is equivalent to $\Delta L \leq0$, as well as
$\lim_{t \to\infty} \widehat{X}_t = \infty$ is equivalent to $\lim_{t
\to
\infty} L_t = 0$. Therefore, in view of Theorem \ref{thm:locmarttoclock},
Theorem \ref{thm:moreimportanceofnumeraire}
will be proved
as long as $L$ is shown to be a nonnegative local martingale. Note that
we already know that $L$ is a supermartingale with $L > 0$ and $L_- >
0$, as follows by the definition of $\widehat{X}$.

Since both $\widehat{X}_- > 0$ and $\widehat{X}> 0$ hold, we have
$\widehat{X}= 1 +
\int
_{(0, \cdot]} \widehat{X}_{t-} \langle\rho_t , dS_t \rangle$ for some
$d$-dimensional predictable and $S$-integrable process $\rho$. A
straightforward\vspace*{1pt} application of Lemma 3.4 in \cite{MR2335830} shows that
$L = 1 - \int_{(0, \cdot]} L_{t-} \langle\rho_t , d\widehat{S}_t
\rangle$, where
\[
\widehat{S}:= S - \biggl[\cS, \int_{(0, \cdot]} \langle\rho_t , d \cS
_t \rangle \biggr] - \sum_{t \leq\cdot} \frac{\Delta\widehat
{X}_t}{\widehat{X}_t} \Delta S_t
\]
with $\cS$ denoting the uniquely defined continuous local martingale
part of $S$ (see, e.g., \cite{MR1943877}).
Since $L_- > 0$ and $L > 0$, $L$ is a local martingale if and only if
$\int_{(0, \cdot]} \langle\rho_t , d\widehat{S}_t \rangle$ is a
local martingale.
The supermartingale property of $L$ already gives that $\int_{(0,
\cdot
]} \langle\rho_t , d\widehat{S}_t \rangle$ is a local
submartingale. We shall show
that $\int_{(0, \cdot]} \langle\rho_t , d\widehat{S}_t \rangle$
is also a local
supermartingale. Since $\langle2 \rho , \Delta S \rangle = 2 (\Delta
\widehat{X}/
\widehat{X}_-) \geq0$, the process $X'$ defined\vspace*{1pt} implicitly via $X' =
1 +
\int
_{(0, \cdot]} X'_{t-} \langle2 \rho_t , dS_t \rangle$ is an element of
$\mathcal{X}$
with $X' > 0$ and $X'_- > 0$. Therefore,\vspace*{2pt} $X' / \widehat{X}$ is a nonnegative
supermartingale. Again, Lemma 3.4 in \cite{MR2335830} shows that $X' /
\widehat{X}= 1 + \int_{(0, \cdot]} (X'_{t-} / \widehat{X}_{t-})
\langle\rho _t , d \widehat{S}_t \rangle$. The supermartingale
property of $X' / \widehat{X}$ implies that
$\int_{(0, \cdot]} \langle\rho_t , d\widehat{S}_t \rangle$ is a local
supermartingale. As $\int_{(0, \cdot]} \langle\rho_t , d\widehat
{S} _t \rangle$ is
a local submartingale, we conclude that $\int_{(0, \cdot]} \langle
\rho _t , d\widehat{S}_t \rangle$ (and, therefore, $L$) is a local
martingale.
\end{pf}

%

%
\printaddresses


\begin{thebibliography}{27}

\bibitem{MR2223957}
%
\begin{barticle}[mr]
\bauthor{\bsnm{Ankirchner},~\bfnm{Stefan}\binits{S.}},
\bauthor{\bsnm{Dereich},~\bfnm{Steffen}\binits{S.}} \AND
\bauthor{\bsnm{Imkeller},~\bfnm{Peter}\binits{P.}}
(\byear{2006}).
\btitle{The {S}hannon information of filtrations and the additional logarithmic
utility of insiders}.
\bjournal{Ann. Probab.}
\bvolume{34}
\bpages{743--778}.
\bid{doi={10.1214/009117905000000648}, mr={2223957}}
\end{barticle}
%
\endbibitem

\bibitem{MR1277002}
%
\begin{barticle}[mr]
\bauthor{\bsnm{Ansel},~\bfnm{Jean-Pascal}\binits{J.-P.}} \AND
\bauthor{\bsnm{Stricker},~\bfnm{Christophe}\binits{C.}}
(\byear{1994}).
\btitle{Couverture des actifs contingents et prix maximum}.
\bjournal{Ann. Inst. H. Poincar\'e Probab. Statist.}
\bvolume{30}
\bpages{303--315}.
\bid{mr={1277002}}
\end{barticle}
%
\endbibitem

\bibitem{MR1849424}
%
\begin{barticle}[mr]
\bauthor{\bsnm{Becherer},~\bfnm{Dirk}\binits{D.}}
(\byear{2001}).
\btitle{The numeraire portfolio for unbounded semimartingales}.
\bjournal{Finance Stoch.}
\bvolume{5}
\bpages{327--341}.
\bid{doi={10.1007/PL00013535}, mr={1849424}}
\end{barticle}
%
\endbibitem

\bibitem{MR0059834}
%
\begin{barticle}[mr]
\bauthor{\bsnm{Bernoulli},~\bfnm{Daniel}\binits{D.}}
(\byear{1954}).
\btitle{Exposition of a new theory on the measurement of risk}.
\bjournal{Econometrica}
\bvolume{22}
\bpages{23--36}.
\bid{mr={0059834}}
\end{barticle}
%
\endbibitem

\bibitem{MR2456470}
%
\begin{barticle}[mr]
\bauthor{\bsnm{Blanchet-Scalliet},~\bfnm{Christophette}\binits{C.}},
\bauthor{\bsnm{El~Karoui},~\bfnm{Nicole}\binits{N.}},
\bauthor{\bsnm{Jeanblanc},~\bfnm{Monique}\binits{M.}} \AND
\bauthor{\bsnm{Martellini},~\bfnm{Lionel}\binits{L.}}
(\byear{2008}).
\btitle{Optimal investment decisions when time-horizon is uncertain}.
\bjournal{J. Math. Econom.}
\bvolume{44}
\bpages{1100--1113}.
\bid{doi={10.1016/j.jmateco.2007.09.004}, mr={2456470}}
\end{barticle}
%
\endbibitem

\bibitem{MR2212119}
%
\begin{barticle}[mr]
\bauthor{\bsnm{Bouchard},~\bfnm{Bruno}\binits{B.}} \AND
\bauthor{\bsnm{Pham},~\bfnm{Huy{\^e}n}\binits{H.}}
(\byear{2004}).
\btitle{Wealth-path dependent utility maximization in incomplete markets}.
\bjournal{Finance Stoch.}
\bvolume{8}
\bpages{579--603}.
\bid{doi={10.1007/s00780-004-0125-8}, mr={2212119}}
\end{barticle}
%
\endbibitem

\bibitem{MR1768009}
%
\begin{bincollection}[mr]
\bauthor{\bsnm{Brannath},~\bfnm{W.}\binits{W.}} \AND
\bauthor{\bsnm{Schachermayer},~\bfnm{W.}\binits{W.}}
(\byear{1999}).
\btitle{A bipolar theorem for {$L^0_+(\Omega,\mathscr F,\mathbf{P})$}}.
In \bbooktitle{S\'eminaire de {P}robabilit\'es, {XXXIII}}.
\bseries{Lecture Notes in Math.}
\bvolume{1709}
\bpages{349--354}.
\bpublisher{Springer}, \baddress{Berlin}.
\bid{doi={10.1007/BFb0096525}, mr={1768009}}
\end{bincollection}
%
\endbibitem

\bibitem{MR1304434}
%
\begin{barticle}[mr]
\bauthor{\bsnm{Delbaen},~\bfnm{Freddy}\binits{F.}} \AND
\bauthor{\bsnm{Schachermayer},~\bfnm{Walter}\binits{W.}}
(\byear{1994}).
\btitle{A general version of the fundamental theorem of asset pricing}.
\bjournal{Math. Ann.}
\bvolume{300}
\bpages{463--520}.
\bid{doi={10.1007/BF01450498}, mr={1304434}}
\end{barticle}
%
\endbibitem

\bibitem{MR1478722}
%
\begin{bincollection}[mr]
\bauthor{\bsnm{Elworthy},~\bfnm{K.~D.}\binits{K.~D.}},
\bauthor{\bsnm{Li},~\bfnm{X.~M.}\binits{X.~M.}} \AND
\bauthor{\bsnm{Yor},~\bfnm{M.}\binits{M.}}
(\byear{1997}).
\btitle{On the tails of the supremum and the quadratic variation of strictly
local martingales}.
In \bbooktitle{S\'eminaire de {P}robabilit\'es, {XXXI}}.
\bseries{Lecture Notes in Math.}
\bvolume{1655}
\bpages{113--125}.
\bpublisher{Springer}, \baddress{Berlin}.
\bid{doi={10.1007/BFb0119298}, mr={1478722}}
\end{bincollection}
%
\endbibitem

\bibitem{MR1970286}
%
\begin{barticle}[mr]
\bauthor{\bsnm{Goll},~\bfnm{Thomas}\binits{T.}} \AND
\bauthor{\bsnm{Kallsen},~\bfnm{Jan}\binits{J.}}
(\byear{2003}).
\btitle{A complete explicit solution to the log-optimal portfolio problem}.
\bjournal{Ann. Appl. Probab.}
\bvolume{13}
\bpages{774--799}.
\bid{doi={10.1214/aoap/1050689603}, mr={1970286}}
\end{barticle}
%
\endbibitem

\bibitem{MR1219534}
%
\begin{bbook}[mr]
\bauthor{\bsnm{He},~\bfnm{Sheng~Wu}\binits{S.~W.}},
\bauthor{\bsnm{Wang},~\bfnm{Jia~Gang}\binits{J.~G.}} \AND
\bauthor{\bsnm{Yan},~\bfnm{Jia~An}\binits{J.~A.}}
(\byear{1992}).
\btitle{Semimartingale Theory and Stochastic Calculus}.
\bpublisher{Kexue Chubanshe (Science Press)}, \baddress{Beijing}.
\bid{mr={1219534}}
\end{bbook}
%
\endbibitem

\bibitem{MR1943877}
%
\begin{bbook}[mr]
\bauthor{\bsnm{Jacod},~\bfnm{Jean}\binits{J.}} \AND
\bauthor{\bsnm{Shiryaev},~\bfnm{Albert~N.}\binits{A.~N.}}
(\byear{2003}).
\btitle{Limit Theorems for Stochastic Processes},
\bedition{2nd} ed.
\bseries{Grundlehren der Mathematischen Wissenschaften [Fundamental Principles
of Mathematical Sciences]}
\bvolume{288}.
\bpublisher{Springer}, \baddress{Berlin}.
\bid{mr={1943877}}
\end{bbook}
%
\endbibitem

\bibitem{MR2335830}
%
\begin{barticle}[mr]
\bauthor{\bsnm{Karatzas},~\bfnm{Ioannis}\binits{I.}} \AND
\bauthor{\bsnm{Kardaras},~\bfnm{Constantinos}\binits{C.}}
(\byear{2007}).
\btitle{The num\'eraire portfolio in semimartingale financial models}.
\bjournal{Finance Stoch.}
\bvolume{11}
\bpages{447--493}.
\bid{doi={10.1007/s00780-007-0047-3}, mr={2335830}}
\end{barticle}
%
\endbibitem

\bibitem{MR0090494}
%
\begin{barticle}[mr]
\bauthor{\bsnm{Kelly},~\bfnm{J.~L.}\binits{J.~L.} \bsuffix{Jr.}}
(\byear{1956}).
\btitle{A new interpretation of information rate}.
\bjournal{Bell. System Tech. J.}
\bvolume{35}
\bpages{917--926}.
\bid{mr={0090494}}
\end{barticle}
%
\endbibitem

\bibitem{RePEcucpjpolecv67y1959p144}
%
\begin{barticle}[vtex]
\bauthor{\bsnm{Latane},~\bfnm{H.~A.}\binits{H.~A.}}
(\byear{1959}).
\btitle{Criteria for choice among risky
ventures}.
\bjournal{Journal of Political Economy}
\textbf{67}
\bpages{144--155}.
\end{barticle}
%
\endbibitem

\bibitem{LONG}
%
\begin{barticle}[auto:SpringerTagBib|2009-01-14|16:51:27]
\bauthor{\bsnm{Long},~\bfnm{J.~B.}\binits{J.~B.} \bsuffix{Jr.}}
(\byear{1990}).
\btitle{The num\'{e}raire portfolio}.
\bjournal{Journal of Financial Economics}
\bvolume{26}
\bpages{29--69}.
\end{barticle}
%
\endbibitem

\bibitem{mascolellwhinstongreen}
%
\begin{bbook}[auto:SpringerTagBib|2009-01-14|16:51:27]
\bauthor{\bsnm{Mas-Colell},~\bfnm{A.}\binits{A.}},
\bauthor{\bsnm{Whinston},~\bfnm{M.~D.}\binits{M.~D.}}
\AND
\bauthor{\bsnm{Green},~\bfnm{J.~R.}\binits{J.~R.}}
(\byear{1995}).
\btitle{Microeconomic Theory}.
\bpublisher{Oxford Univ. Press},
\baddress{Oxford}.
\end{bbook}
%
\endbibitem

\bibitem{MR2247846}
%
\begin{barticle}[mr]
\bauthor{\bsnm{Nikeghbali},~\bfnm{Ashkan}\binits{A.}} \AND
\bauthor{\bsnm{Yor},~\bfnm{Marc}\binits{M.}}
(\byear{2006}).
\btitle{Doob's maximal identity, multiplicative decompositions and enlargements
of filtrations}.
\bjournal{Illinois J. Math.}
\bvolume{50}
\bpages{791--814}.
\bid{mr={2247846}}
\end{barticle}
%
\endbibitem

\bibitem{MR2267213}
%
\begin{bbook}[mr]
\bauthor{\bsnm{Platen},~\bfnm{Eckhard}\binits{E.}} \AND
\bauthor{\bsnm{Heath},~\bfnm{David}\binits{D.}}
(\byear{2006}).
\btitle{A Benchmark Approach to Quantitative Finance}.
\bpublisher{Springer}, \baddress{Berlin}.
\bid{doi={10.1007/978-3-540-47856-0}, mr={2267213}}
\end{bbook}
%
\endbibitem

\bibitem{MR1780932}
%
\begin{bbook}[mr]
\bauthor{\bsnm{Rogers},~\bfnm{L.~C.~G.}\binits{L.~C.~G.}} \AND
\bauthor{\bsnm{Williams},~\bfnm{David}\binits{D.}}
(\byear{2000}).
\btitle{Diffusions, {M}arkov Processes, and Martingales}.
\bseries{Cambridge Mathematical Library}
\bvolume{2}.
\bpublisher{Cambridge Univ. Press}, \baddress{Cambridge}.
\bid{mr={1780932}}
\end{bbook}
%
\endbibitem

\bibitem{MR0295739}
%
\begin{barticle}[mr]
\bauthor{\bsnm{Samuelson},~\bfnm{Paul~A.}\binits{P.~A.}}
(\byear{1971}).
\btitle{The ``fallacy'' of maximizing the geometric mean in long
sequences of
investing or gambling}.
\bjournal{Proc. Natl. Acad. Sci. USA}
\bvolume{68}
\bpages{2493--2496}.
\bid{mr={0295739}}
\end{barticle}
%
\endbibitem

\bibitem{RePEceeejbfinav3y1979i4p305307}
%
\begin{barticle}[auto:SpringerTagBib|2009-01-14|16:51:27]
\bauthor{\bsnm{Samuelson},~\bfnm{Paul~A.}\binits{P.~A.}}
(\byear{1979}).
\btitle{Why we should not make mean log of wealth big though years to
act are long}.
\bjournal{Journal of Banking and Finance}
\bvolume{3}
\bpages{305--307}.
\end{barticle}
%
\endbibitem

\bibitem{MR0348870}
%
\begin{bbook}[mr]
\bauthor{\bsnm{Savage},~\bfnm{Leonard~J.}\binits{L.~J.}}
(\byear{1972}).
\btitle{The Foundations of Statistics}, \bedition{revised} ed.
\bpublisher{Dover}, \baddress{New York}.
\bid{mr={0348870}}%
\end{bbook}%
%
\endbibitem%

\bibitem{MR2316805}
%
\begin{bbook}[mr]
\bauthor{\bparticle{von }\bsnm{Neumann},~\bfnm{John}\binits{J.}}
\AND
\bauthor{\bsnm{Morgenstern},~\bfnm{Oskar}\binits{O.}}
(\byear{2007}).
\btitle{Theory of Games and Economic Behavior}, \bedition
{anniversary} ed.
\bpublisher{Princeton Univ. Press}, \baddress{Princeton, NJ}.
\bid{mr={2316805}}
\end{bbook}
%
\endbibitem

\bibitem{Williams36}
%
\begin{barticle}[vtex]
\bauthor{\bsnm{Williams},~\bfnm{J.~B.}\binits{J.~B.}}
(\byear{1936}).
\btitle{Speculation and the carryover}.
\bjournal{Quarterly Journal of Economics}
\bvolume{50}
\bpages{436--455}.
\end{barticle}
%
\endbibitem

\bibitem{MR2114989}
%
\begin{barticle}[mr]
\bauthor{\bsnm{{\v{Z}}itkovi{\'c}},~\bfnm{Gordan}\binits{G.}}
(\byear{2005}).
\btitle{Utility maximization with a stochastic clock and an unbounded random
endowment}.
\bjournal{Ann. Appl. Probab.}
\bvolume{15}
\bpages{748--777}.
\bid{doi={10.1214/105051604000000738}, mr={2114989}}
\end{barticle}
%
\endbibitem

\bibitem{Zit08}
%
\begin{barticle}[auto:SpringerTagBib|2009-01-14|16:51:27]
\bauthor{\bsnm{{\v{Z}}itkovi{\'c}},~\bfnm{Gordan}\binits{G.}}
(\byear{2008}).
\btitle{Convex-compactness and its applications}.
\bjournal{Math. Financ. Econ.}
\bmisc{To appear.}
\end{barticle}
%
\endbibitem

\end{thebibliography}
\end{document}